\documentclass[prb,aps,twocolumn,superscriptaddress, floatfix]{revtex4-1}

\usepackage{graphicx}% Include figure files
\usepackage{xcolor}
\usepackage{soul}
\usepackage{dcolumn}
\usepackage{bm}
\usepackage{amsmath}
\usepackage{amssymb}

\usepackage{ulem}

\usepackage[right=2.0cm, left=2.0cm, bottom=2.5cm, top=2.5cm]{geometry}

\usepackage{url}

\usepackage{hyperref}

\hypersetup{colorlinks=true,
    linkcolor=purple,
   citecolor=purple, urlcolor=purple, breaklinks=true}

\begin{document}

\title{Towards reliable synthesis of superconducting infinite layer nickelate thin films by topochemical reduction}

\author{Araceli \surname{Guti\'{e}rrez--Llorente}}
\email[]{araceli.gutierrez@urjc.es}
%%\altaffiliation[]{xxx}
\affiliation{Universidad Rey Juan Carlos, Escuela Superior de Ciencias Experimentales y Tecnolog\'{i}a, Madrid 28933, Spain}
\affiliation{Laboratoire Albert Fert CNRS, Thales, Universit\'{e} Paris Saclay, 91767 Palaiseau, France}

\author{Aravind Raji}
\affiliation{Universit\'{e} Paris Saclay, CNRS, Laboratoire de Physique des Solides, 91405 Orsay, France}
\affiliation{Synchrotron SOLEIL, L’Orme des Merisiers, BP 48 St Aubin, Gif sur Yvette, 91192, France}

\author{Dongxin Zhang}
\affiliation{Laboratoire Albert Fert CNRS, Thales, Universit\'{e} Paris Saclay, 91767 Palaiseau, France}

\author{Laurent Divay}
\affiliation{Thales Research \& Technology France, 91767 Palaiseau, France}

\author{Alexandre Gloter}
\affiliation{Universit\'{e} Paris Saclay, CNRS, Laboratoire de Physique des Solides, 91405 Orsay, France}

\author{Fernando Gallego}
\affiliation{Laboratoire Albert Fert CNRS, Thales, Universit\'{e} Paris Saclay, 91767 Palaiseau, France}

\author{Christophe Galindo}
\affiliation{Thales Research \& Technology France, 91767 Palaiseau, France}

\author{Manuel Bibes}
\affiliation{Laboratoire Albert Fert CNRS, Thales, Universit\'{e} Paris Saclay, 91767 Palaiseau, France}

\author{Luc\'{i}a Iglesias}
\email[]{lucia.iglesias@cnrs-thales.fr}
\affiliation{Laboratoire Albert Fert CNRS, Thales, Universit\'{e} Paris Saclay, 91767 Palaiseau, France}

%\date{\today}

\begin{abstract}

Infinite layer nickelates provide a new route beyond copper oxides to address outstanding questions in the field of unconventional superconductivity. However, their synthesis poses considerable challenges, largely hindering experimental research on this new class of oxide superconductors. That synthesis is achieved in a two-step process that yields the most thermodynamically stable perovskite phase first, then the infinite-layer phase by topotactic reduction, the quality of the starting phase playing a crucial role.  Here, we report on reliable synthesis of superconducting infinite-layer nickelate films after successive topochemical reductions of a parent perovskite phase with nearly optimal stoichiometry.  Careful analysis of the transport properties of the incompletely reduced films reveals an improvement of the strange metal behaviour of their normal state resistivity over subsequent topochemical reductions, offering insight into the reduction process.

\end{abstract}

\maketitle %\maketitle must follow title, authors, abstract and \pacs

\section{\label{sec:intro}Introduction}

The first observation of superconductivity at relatively high temperature in single-crystal thin films of infinite-layer NdNiO$_2$ upon hole doping was a significant breakthrough.\cite{Li:19}  Thereafter, superconductivity has also been observed in other families of hole-doped infinite-layer (IL) nickelate thin films, such as (Pr, Sr)NiO$_2$,\cite{osada:20a,osada:20b,wang:22} (La, Sr)NiO$_2$,\cite{osada:21,osada:23} (La, Ca)NiO$_2$,\cite{zeng:22} and (Nd, Eu)NiO$_2$;\cite{wei:23} in reduced Ruddlesden-Popper  Nd$_6$Ni$_5$O$_{12}$ thin films without chemical doping;\cite{pan:21} and in bilayer Ruddlesden–Popper La$_3$Ni$_2$O$_7$ bulk single-crystals under high pressure.\cite{sun:23}  Thus, this new class of oxide superconductors provides a new route beyond copper oxides to address outstanding questions in the field of unconventional superconductivity,\cite{zhou:21} such as the mechanism that causes the electrons to form pairs, unanswered despite decades of intense research activity.\cite{scalapino:12,tsuei:00,keimer:15,proust:19,stewart:17,O_Mahonya:22}

Despite an unremarkable critical temperature of around 10 K, that initial observation generated intense interest.\cite{pickett:21,norman:20,botana:22,mitchell:21}  Part of the reason for that is the discovery of superconducting (SC) nickelates was driven by the decades-long search of cuprate-like physics in other strongly correlated metallic oxides. Within this context, Ni-based compounds with Ni ions arranged in corner-sharing NiO$_4$ square units and ultralow chemical valence (Ni$^{1+}$ and 3d$^9$ configuration) were suggested to exhibit superconductivity due to their electronic and structural similarities with the Cu$^{2+}$ ions in cuprates.\cite{Anisimov:99}  However, besides these similarities, very different behaviour between LaNiO$_2$ and CaCuO$_2$ was pointed out early on.\cite{lee:04}  And, whether there is a universal mechanism that explains superconductivity in both cuprates and nickelates remains an open question.\cite{botana:20,nomura:22,gu:22,kitatani:23,hepting:21,lechermann:20,rossi:21,goodge:21,carrasco-alvarez:22}

Further progress in the field crucially depends on the synthesis of high quality SC nickelate samples, which can provide reliable experimental data.  However, only a few groups worldwide have developed the appropriate expertise to date,\cite{Li:19,wang:22,zeng:22,wei:23} since the synthesis of these films poses serious challenges.

Epitaxial growth of complex oxides thin films, such as nickelates, is achieved under vacuum at high substrate temperatures to increase surface mobility of adatoms and improve crystallinity of the grown films.  At these elevated temperatures, the process yields the most thermodynamically stable phase with octahedral NiO$_6$ coordination.  Thus, the synthesis of the reduced structure with square-planar coordination around Ni$^{1+}$ ions, arranged in infinite layers, should proceed via the subsequent removal of relatively mobile oxygen anions at low temperatures, by means of kinetically controlled reactions that enable the preparation of metastable phases.\cite{hayward:07}

\begin{figure*}[!htb]
 \includegraphics[keepaspectratio=true, width=\linewidth]{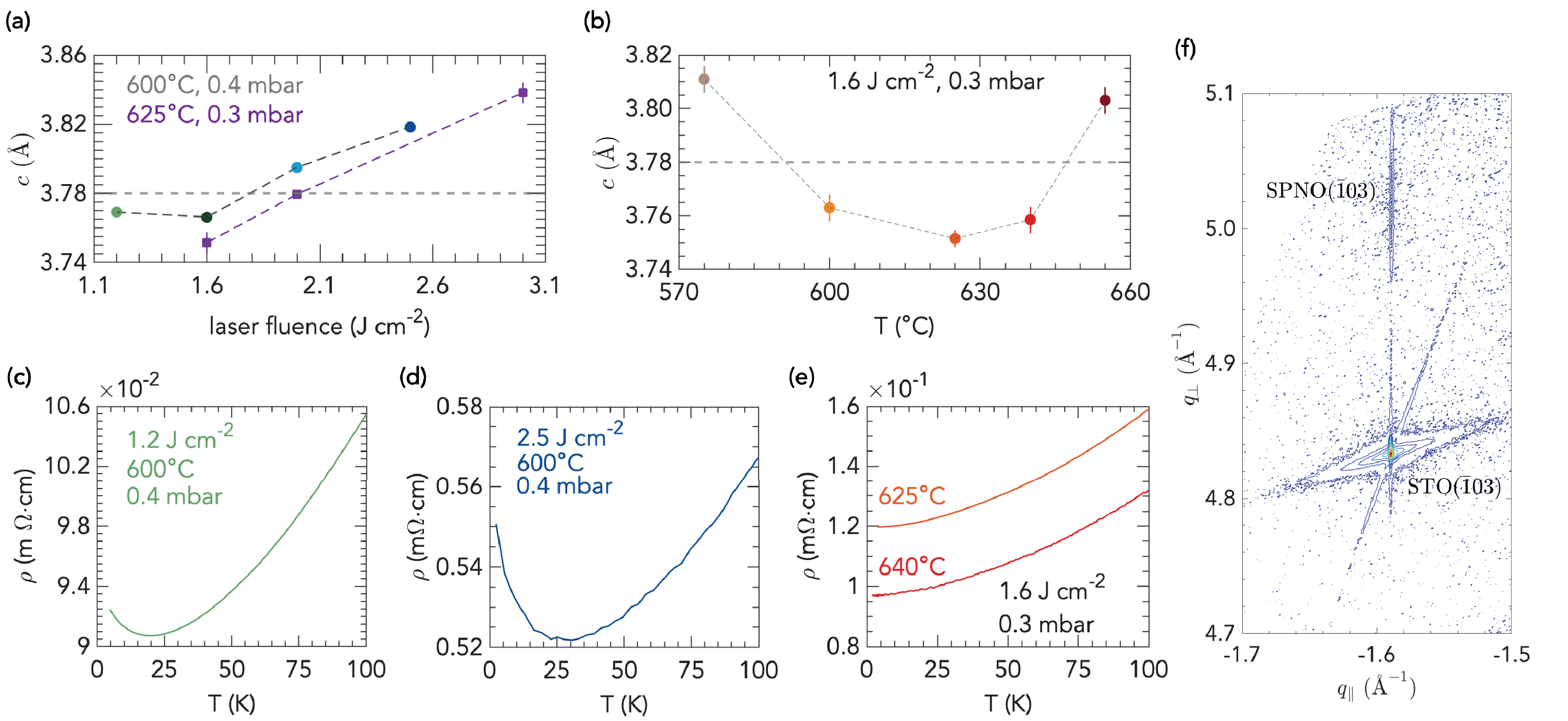} \caption{\label{Fig_01} {{\bf{Optimization of growth conditions for PSNO$_3$.}}  {\bf{(a)}} $c$-axis lattice parameter of PSNO$_3$ films as function of laser fluence at various conditions of substrate temperature and oxygen pressure in the PLD chamber during growth. {\bf{(b)}} $c$-axis lattice parameter as function of substrate temperature, keeping constant laser fluence at 1.6 Jcm$^{-2}$ and oxygen pressure at 0.3 mbar.  The dashed horizontal line in panels (a) and (b) is an estimate of $c$ from the bulk lattice constant of PNO$_3$ strained to the STO substrate, and doping with Sr is expected to bring about a contraction of the unit cell.  {\bf{(c, d, e)}} Temperature dependence of resistivity, $\rho(T)$, in PSNO$_3$ films grown at different conditions of laser fluence, substrate temperature and oxygen pressure, whose $c$-axis lattice parameters are shown in panel (a) or (b).  {\bf{(f)}} High-resolution RSMs around the $(\bar{1}03)$ reflection of a PSNO$_3$ film under optimized conditions (1.6 $\mbox{J}\,\mbox{cm}^{-2}$, 640 $^{\circ}$C, 0.3 mbar); indices with respect to the pseudocubic unit cell. The ($\bar{1}$03) reflection from the STO substrate is also shown.  Error bars in panels ({\textit{a}}) and ({\textit{b}}) indicate the $1\sigma$ uncertainties of the fits.
}}
\end{figure*}

This transformation of the starting complex oxide phases into oxygen-deficient metastable phases can be triggered by different methods,\cite{ranmohotti:11,meng:23,kageyama:18} and is topotactic since it does not involve diffusive rearrangement of the host cations, although lattice parameters and bond lengths change, giving rise to drastic changes in electronic structures.  Thus far, in the case of hole-doped nickelates, the topotactic transition from the perovskite phase into the highly metastable SC IL phase has been accomplished {\textit{ex-situ}} by low temperature annealing within a sealed glass ampoule using CaH$_2$ as the reducing agent,\cite{Li:19,osada:20a,osada:21} and {\textit{in-situ}} using an oxygen getter metal layer.\cite{wei:23}  Under the former topochemical approach, the resulting phase can be tuned by the choice of the metal hydride and by the reaction conditions, in particular, temperature and time.\cite{hayward:99,hayward:03,kawai:10}  On the one hand, the temperature of the topochemical reaction has to be high enough to bring about the reduction as the activity of metal hydrides in solid state reduction declines at lower temperature.  On the other hand, the perovskite framework is more stable at low temperature since an increase in temperature can result in non-topotactic reactions, if the cations in the resulting metastable phase become mobile, leading to degradation of the sample crystallinity. Furthermore, the crystalline quality of the starting perovskite phase greatly affects the reaction.\cite{lee:20}  Indeed, decreasing the lattice mismatch (tensile strain) between the parent perovskite phase and the substrate enhances crystallinity of the subsequent reduced phase, even though the increase in the in-plane lattice parameter upon reduction leads to higher compressive strain for the reduced phase in that case.\cite{lee:23}

Here, we report the successful synthesis of SC strontium-doped praseodymium nickelate thin films.  Complementary to previous approaches, we study the cation stoichiometry of the starting perovskite films by X-ray photoelectron spectroscopy.  Furthermore, we carry out a comprehensive study on transport properties of intermediate reduced films, and find an enhancement of the strange metal behaviour of the normal state resistivity of incompletely reduced films over subsequent topochemical reductions.  Moreover, the removal of apical oxygen anions from the perovskite phase is confirmed through the structural analysis of the IL phase using four dimensional scanning transmission electron microscopy (4D-STEM).

\section{\label{sec:results}Results and discussion}

As a first step, we optimize the growth conditions of the Pr$_{0.8}$Sr$_{0.2}$NiO$_3$ parent perovskite films (hereafter, PSNO$_3$) with thicknesses between 10 and 13 unit cells (u.c.) by pulsed laser deposition (PLD) on TiO$_2$-terminated (001)-oriented SrTiO$_3$ (STO) substrates.  Bulk PrNiO$_3$ crystallizes in an orthorhombic structure (space group \#62  \textit{Pbnm}, GdFeO$_3$-type) with lattice constants at room temperature  of $a=5.42\;\mbox{\AA}$, $b=5.38 \; \mbox{\AA}$, $c=7.63 \; \mbox{\AA}$ (pseudocubic constant $a_{pc} \approx 3.82 \; \mbox{\AA}$).\cite{garcia-munoz:92}  Its crystalline structure is not modified by doping with Sr$^{2+}$ although this brings about a contraction of the unit cell, despite the larger effective size of Sr$^{2+}$ compared to Pr$^{3+}$.\cite{garcia-munoz:95}  Thus, assuming a Poisson ratio of $\nu=0.3$, a value common to other oxide perovskites,\cite{ledbetter:90} the expected out-of-plane lattice parameter for epitaxially grown PSNO$_3$ films on STO substrates, inducing 2.23\% of tensile strain, is $\lesssim 3.78 \; \mbox{\AA}$.  This estimate leads to a value of 2$\theta\gtrsim 48.04^{\circ}$ for the $(002)_{pc}$ reflexion in the x-ray diffraction pattern (using Copper K-$\alpha$ radiation).  Interestingly, that value is in agreement with the threshold experimentally found in the (Nd, Sr)NiO$_3$ system, where if the (002) pseudocubic perovskite peak $2\theta$ position is below $\approx 48^{\circ}$, the subsequently reduced film never exhibits superconductivity.\cite{lee:20}.

%Pr$_{0.8}$Sr$_{0.2}$NiO$_3$
We use complementary information from X-ray Diffraction (XRD) and resistivity as function of temperature, $\rho(T)$, to elucidate the optimal growth conditions.  Laser fluence has a major impact on the film quality, as expected.\cite{ohnishi:05,ohnishi:08}  Initially, we have explored laser fluences ranging from 1.2 $\mbox{J}\,\mbox{cm}^{-2}$ to 3 $\mbox{J}\,\mbox{cm}^{-2}$ at temperatures around 600 $^{\circ}$C in a strongly oxidizing environment with 0.3-0.4 mbar of O$_2$, required to stabilize a Ni$^{3.2+}$ oxidation state.

As illustrated in Fig.\ref{Fig_01}(a), the out-of-plane lattice parameter $c$ of the PSNO$_3$ films shows a minimum for a laser fluence of 1.6 $\mbox{J}\,\mbox{cm}^{-2}$ for different conditions of oxygen pressure and substrate temperature.  The cell expansion observed as the fluence departs from that value is indicative of cation vacancies in the films. Oxygen vacancies can be ruled out as the cause of the increase in lattice parameter, since the higher the oxygen pressure is, the more the unit cell expands.  A more detailed exploration of the out-of-plane lattice parameter as a function of the substrate temperature at a laser fluence of 1.6 $\mbox{J}\,\mbox{cm}^{-2}$ in 0.3 mbar of O$_2$ is shown in Fig.\ref{Fig_01}(b). Under these conditions, $c$ lattice constants consistent with a low density of defects are found for substrate temperatures in the range from 600$^{\circ}$C to 640$^{\circ}$C.  We confirm the superior quality of the films grown at 1.6 $\mbox{J}\,\mbox{cm}^{-2}$ by electrical transport measurements, as depicted in Fig.\ref{Fig_01}(c, d, e).

Indeed, a resistivity upturn at low temperature is observed at fluences of 1.2 $\mbox{J}\,\mbox{cm}^{-2}$ or 2.5 $\mbox{J}\,\mbox{cm}^{-2}$, with minima at around 19 K and 30 K, respectively, Fig.\ref{Fig_01}(c, d), that may be attributed to localization phenomena either through weak localization or by electron-electron enhanced interactions,\cite{lee:85,bergmann:84} while a metallic state is observed down to 2 K in films grown under 1.6 $\mbox{J}\,\mbox{cm}^{-2}$, 0.3 mbar, and 625$^{\circ}$C or 640$^{\circ}$C, Fig.\ref{Fig_01}(e).  Moreover, reciprocal space maps (RSM) around the asymmetric $(\bar{1}03)$ reflection of PSNO$_3$ films grown under the latter conditions demonstrates that the films are fully strained to the STO substrate, as depicted in Fig.\ref{Fig_01}(f).  See also Fig. S1 for supplementary information on the optimal growth conditions.  Detailed analysis of the resistivity of those films in the low temperature region in the framework of a model that includes quantum corrections is presented in Fig. S2, Supplementary Information.

\begin{figure}[htb]
 \includegraphics[keepaspectratio=true, width=0.8\linewidth]{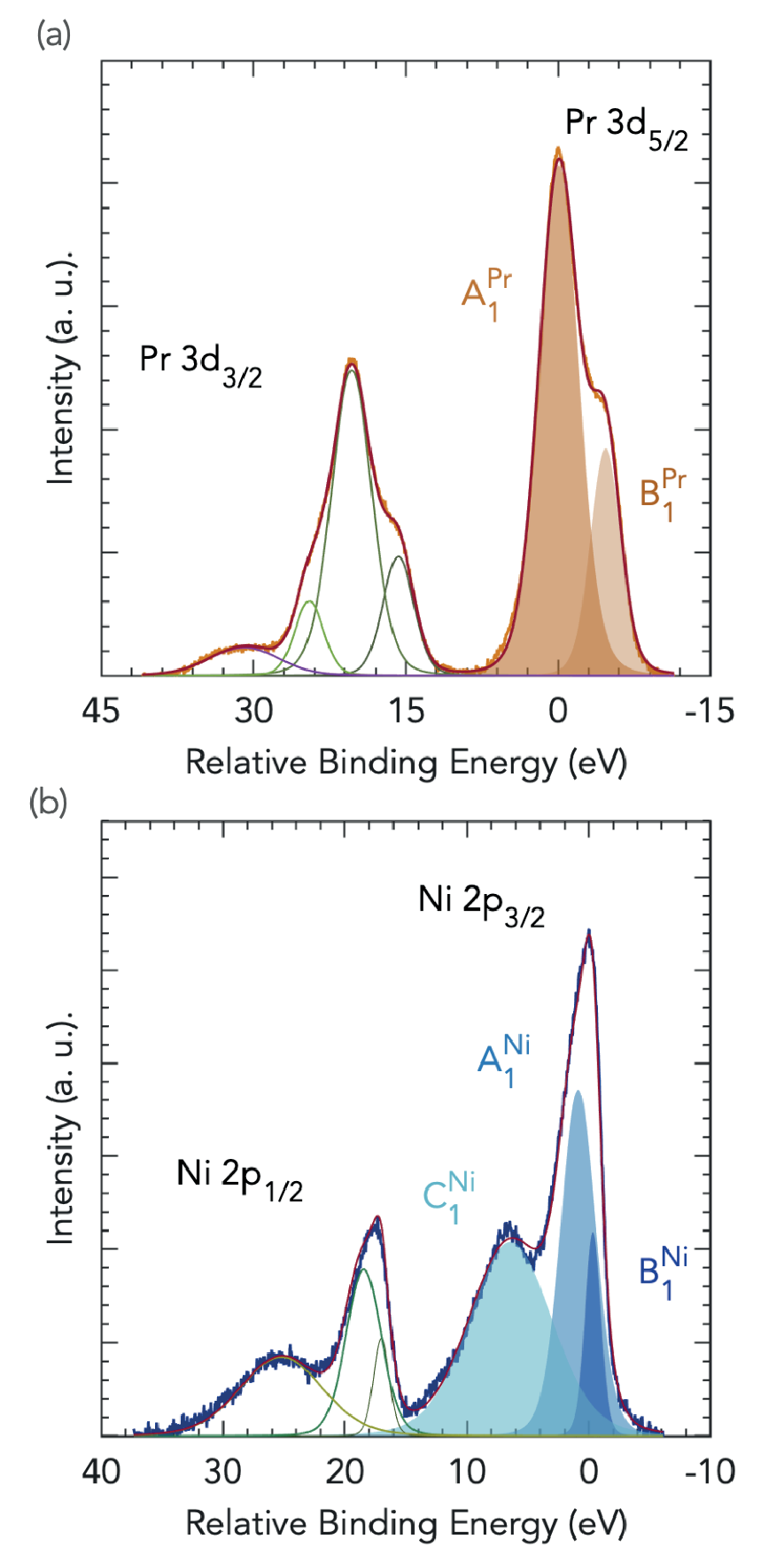} \caption{\label{Fig_XPS_peaks}  {{\bf{Cation stoichiometry of PSNO$_3$ films by XPS.}}  Peak models used for the quantification of the $\mbox{[Pr]/[Ni]}$ ratio. The components ((a) brown areas, Pr; (b) blue areas, Ni) and total fit envelope are shown.
}}
\end{figure}

\begin{figure}[htb]
 \includegraphics[keepaspectratio=true, width=0.7\linewidth]{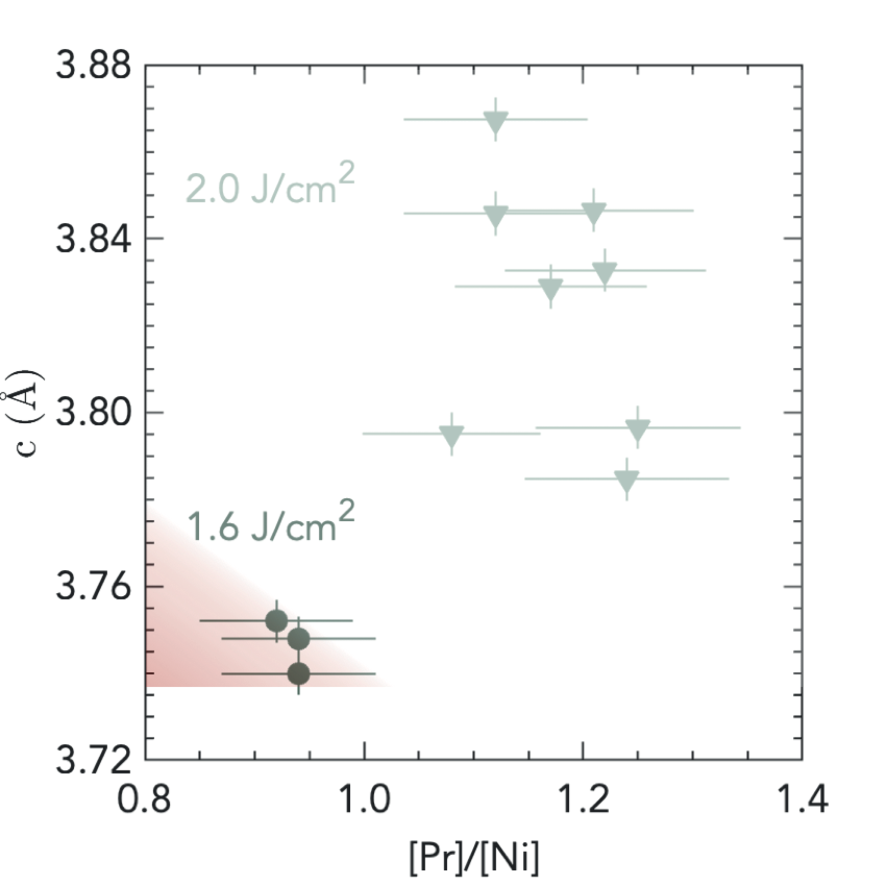} \caption{\label{Fig_XPS_ratios}  {{\bf{$c$-axis lattice parameter of PSNO$_3$ films as a function of cation stoichiometry.}}  $c$ lattice parameter \textit{vs} $\mbox{[Pr]/[Ni]}$ ratio of PSNO$_3$ films grown at laser fluence 1.6 $\mbox{J}\,\mbox{cm}^{-2}$, 0.3 mbar and substrate temperatures of 625 $^{\circ}$C or 640 $^{\circ}$C or under non optimal growth conditions of laser fluence 2 $\mbox{J}\,\mbox{cm}^{-2}$ for different series of films, extracted from the fitted XPS spectra. See Fig.S5 for supplementary information on different growth conditions.  A laser fluence of 1.2 $\mbox{J}\,\mbox{cm}^{-2}$ gives rise to $c=3.77 \;\mbox{\AA}$ and $\mbox{[Pr]/[Ni]}=1.07$.  Error bars of 15\% are applied to the $\mbox{[Pr]/[Ni}$ ratio, which is the expected accuracy of XPS quantification for transition metal oxides.\cite{brundle:20}  The red shaded area represents the target region within the error for optimized films.  
}}
\end{figure}

We probed cation stoichiometry of the PSNO$_3$ films by means of X-ray photoelectron spectroscopy (XPS).  Elemental composition of the films and assignments of the peaks in a XPS survey spectrum are plotted in Fig. S3 (Supplementary Information).  Quantitative XPS for the determination of the $\mbox{[Pr]/[Ni]}$ ratio was derived from the area under the core-level and satellite peaks, Pr 3\textit{d} and Ni 2\textit{p}, as shown in Fig.\ref{Fig_XPS_peaks}, and relative sensitivity factors derived from photoionization cross-sections by Scofield.\cite{scofield:73}.  The precise details regarding the quantification of cation stoichiometry of PSNO$_3$ films are given in Supplementary Information, section 2 and Fig. S4.

Fig.\ref{Fig_XPS_ratios} shows $c$-axis lattice parameter as a function of $\mbox{[Pr]/[Ni]}$ ratio of PSNO$_3$ films grown under optimal growth conditions of substrate temperature and oxygen pressure and a laser fluence of 1.6 $\mbox{J}\,\mbox{cm}^{-2}$, or at a laser fluence of 2 $\mbox{J}\,\mbox{cm}^{-2}$.  Given the complex satellite structure of the XPS spectra, one may expect $\approx 15\%$ accuracy for the XPS quantification.\cite{brundle:20}  Within these limits of accuracy, nearly stoichiometric films are obtained for an optimal laser fluence of 1.6 $\mbox{J}\,\mbox{cm}^{-2}$, while films grown at a laser fluence of 2 Jcm$^{-2}$ present a dramatic increase in the $\mbox{[Pr]/[Ni]}$ ratio, with strong deviations from the expected film stoichiometry.  Optimal cation stoichiometry yields the lowest unit-cell volume, in agreement with previous results about films grown by molecular beam epitaxy.\cite{li:21}  Fig. S5 (Supplementary Information) shows in detail the dependence of $\mbox{[Pr]/[Ni]}$ ratio on oxygen pressure and substrate temperature at 2 $\mbox{J}\,\mbox{cm}^{-2}$.  We also assessed the $\mbox{[Pr]/[Sr]}$ ratio from Pr 3$d_{5/2}$ and Sr 3$d_{5/2}$, and found values close to $\mbox{[Pr]/[Sr]} \approx 4$, indicating no strong deviations relative to the nominal hole-doping levels across different growth conditions, as depicted in Fig. S6 (Supplementary Information).

\begin{figure}[!htb]
 \includegraphics[keepaspectratio=true, width=0.75\linewidth]{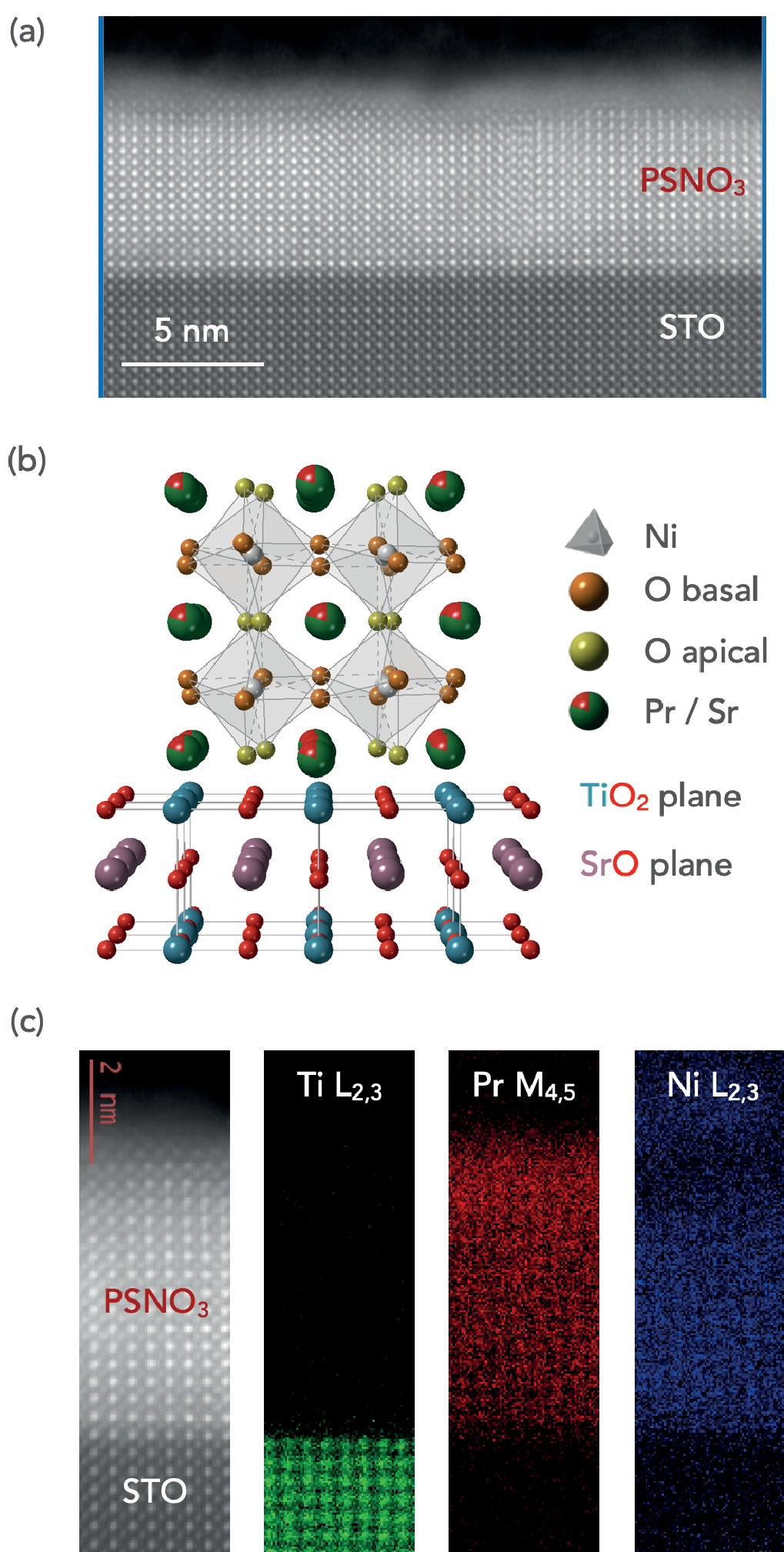} \caption{\label{Fig_TEM_perovskite} {{\bf{Structure of a PSNO$_3$ film grown under optimized conditions on an STO substrate.}}  {\bf{(a)}} HAADF-STEM image of a PSNO$_3$ film grown at a laser fluence of 1.6 $\mbox{J}\,\mbox{cm}^{-2}$, 0.3 mbar, and 625 $^{\circ}$C).  {\bf{(b)}} Schematic of the structural model of the films grown on STO substrate. {\bf{(c)}} Atomic-resolution HAADF-STEM image and simultaneously recorded elemental EELS maps of Ti L$_{2, 3}$, Pr M$_{4, 5}$ and Ni L$_{2, 3}$ edges.
}}
\end{figure}

To provide further structural characterization of the grown films and visualize possible defects of the atomic lattice, cross-sectional scanning transmission electron microscopy (STEM) imaging was carried out.  Fig.\ref{Fig_TEM_perovskite}(a) depicts a high-angle annular dark-field STEM (HAADF-STEM) image of an optimized PSNO$_3$ film over a wide area of the film with very few vertical Ruddlesden–Popper faults (the structural model is illustrated in Fig.\ref{Fig_TEM_perovskite}(b)).  Electron energy-loss spectroscopy (EELS) map analysis indicate a uniform distribution of Ni and Pr across the film and an abrupt interface with the STO substrate, Fig.\ref{Fig_TEM_perovskite}(c)).

\begin{figure*}[!htb]
 \includegraphics[keepaspectratio=true, width=\linewidth]{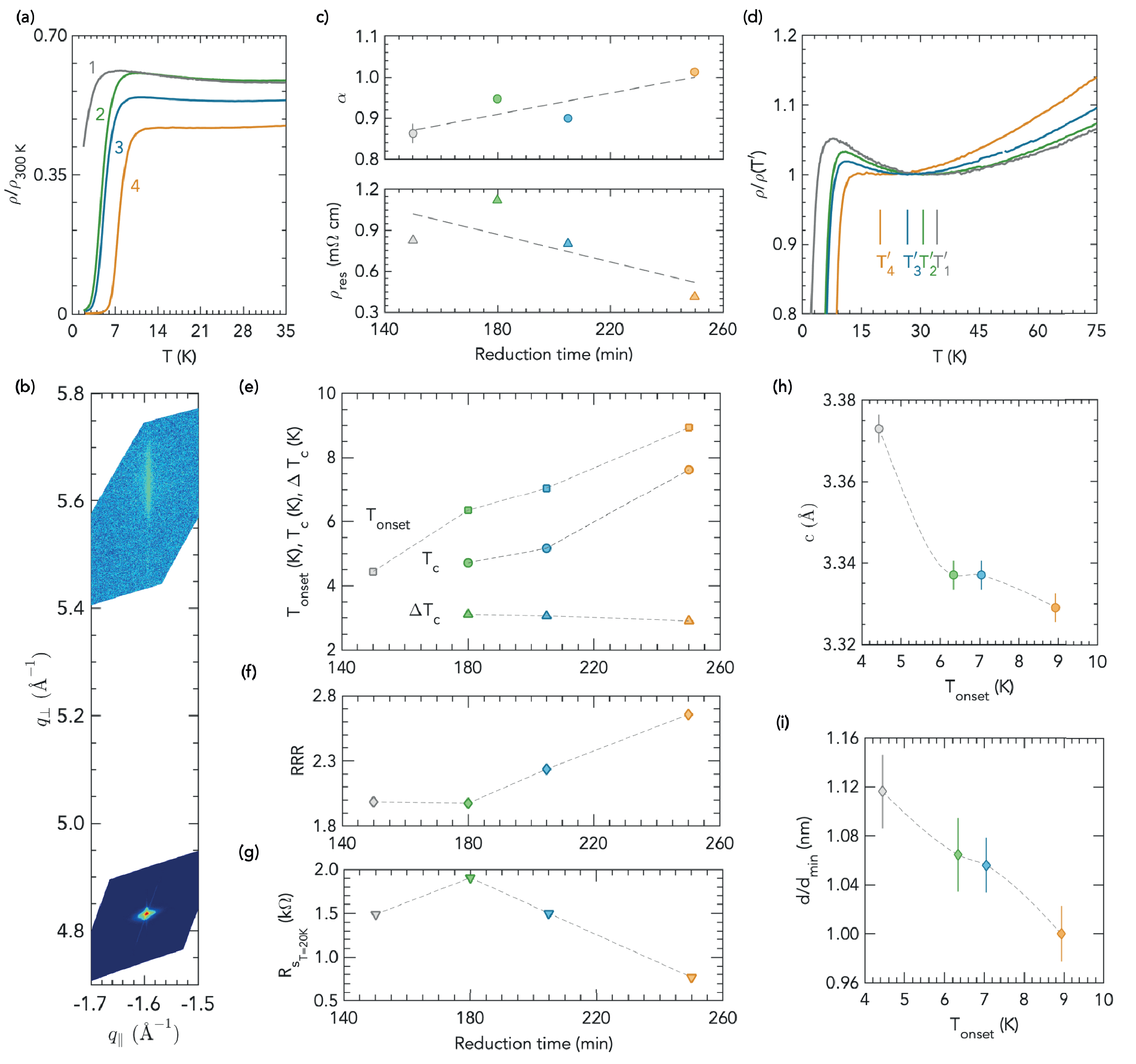} \caption{\label{Fig_03} { {\bf{Topotactic phase transition over subsequent topochemical reductions: Electrical transport, superconducting properties, and structural characterization of reduced films.}} {\bf{(a)}} Temperature dependent resistivity normalised to its room temperature value of a representative PSNO$_2$ film after successive reduction processes carried out at 260$^{\circ}$C for periods of (1) 150 min, (2) 30 min, (3) 25 min, and (4) 45 min.  {\bf{(b)}}  RSM of the complete reduced IL film around the $(\bar{1}03)$ reflection.  {\bf{(c)}} Exponent $\alpha$ (top) and residual resistance $\rho_{res}$ (bottom) in $\rho(T)=\rho_{res}+AT^{\alpha}$.   {\bf{(d)}} Evolution of resistivity, normalised to its local minimum value, depicting upturns at temperatures above the onset of the SC transition.  {\bf{(e)}} Evolution of SC properties as a function of reduction time: onset SC transition temperature, $T_{onset}$, defined as the intersection of linear extrapolations from the normal state and the superconducting transition region; critical temperature, $T_c$, 50\% of normal state at 20 K; and, transition width, $\Delta T_c$, 90\%–10\% of normal state at 20 K.  {\bf{(f)}} Residual Resistivity Ratio (RRR), defined as the resistivity at 300 K divided by the extrapolation of the temperature-linear resistivity to 0 K.  {\bf{(g)}} Sheet resistance $R_s$ at 20 K. {\bf{(h)}} $c$-axis lattice parameter and {\bf{(i)}} thickness of the PSNO$_2$ films normalized to the maximum reduced thickness,  against $T_{onset}$.  Error bars denote the 95\% confidence interval for each fitting (panels {\textit{c}}, {\textit{h}} and {\textit{i}}), and are sometimes smaller than the marker size.  See also Experimental section for details on lattice parameters calculation.  The dashed lines in all the panels are a guide to the eye.
}}
\end{figure*}

Having accomplished the growth of PSNO$_3$ films with close to optimal stoichiometry and minimal defect densities, we now focus on the study of reduced samples Pr$_{0.8}$Sr$_{0.2}$NiO$_2$ (hereafter, PSNO$_2$).  The as-grown 5 $\times$ 5 mm$^2$ films were cut into two pieces with lateral dimensions of 2.5 $\times$ 5 mm$^2$ or four pieces of 2.5 $\times$ 2.5 mm$^2$, approximately, before carrying out the hydride reduction. A same piece from the as-grown sample is repeatedly reduced.  Between annealings, the sealed, evacuated glass tube that contains the reducing agent (CaH$_2$, physically separated from the sample) is unsealed to measure the sample, and, hence, a new tube is evacuated and sealed to carry out the next step. 

\clearpage

Fig.\ref{Fig_03}(a) shows temperature dependent resistivity normalised to its room temperature value (to remove the influence of geometric factors) of a SC PSNO$_2$ film over successive reductions carried out at 260$^{\circ}$C for periods of 150 min (1), 30 min (2), 25 min (3), and 45 min (4), (see Fig. S7, Supplementary Information for the extended range 0 K - 300 K).  Although superconductivity arises after the first reduction, the zero-resistance state is achieved as a result of further heating periods.  The absolute value of resistivity is $\rho(20\;K) \approx 0.4$ $\mbox{m}\Omega\;\mbox{cm}$ for the fully reduced sample (Fig. S8, supplementary information), matching the previously published value for SC PSNO$_2$ films on STO.\cite{osada:20b}  No diffraction reflections of the parent perovskite phase are detected in the XRD patterns of the reduced samples, and positions of the observed reflections are consistent with the (001) and (002) diffractions of the tetragonal IL phase (Fig. S9, Supplementary Information). In addition, the RSM in Fig.\ref{Fig_03}(b) reveals that after the fourth step the reduced phase remains fully strained to the substrate.  This is noteworthy because perovskite PSNO$_3$ thin films on STO substrates undergo a change from tensile strain to compressive strain as they turn into the IL tetragonal PSNO$_2$ phase.  Indeed, upon deintercalation of apical oxygen atoms, the in-plane (out-of-plane) lattice parameter of the IL PSNO$_2$ phase expands (drastically shrinks) relative to that of the parent perovskite PSNO$_3$ phase,\cite{hayward:99,hayward:03} and STO substrates induce a compressive strain ($\approx -1.2 \%$) on that reduced IL phase.

Resistivity of the SC fully reduced film decreases linearly in $T$ from 300 K down to around 60 K, and deviates from linearity below that temperature, likely because of scattering due to disorder, as we discuss below.  $T$-linear normal-state resistivity has been observed in other families of unconventional superconductors, in particular in the high $T$ and low $T$ (superconductivity supressed with a magnetic field) normal-state of cuprates,\cite{taillefer:10,legros:19} and has also been recently found in optimally doped (Nd, Sr)NiO$_2$ films with improved crystallinity and reduced disorder synthesised on LSAT substrates.\cite{lee:23}

We carried out least-square fitting of the experimental data to $\rho(T) = \rho_{res}+AT^{\alpha}$ in the range from 300 K to 60 K to determine the power law dependence of resistivity (Fig. S10, Supplementary Information), and found that linearity within this region increases over subsequent reductions, {\textit{i.e.}} $\alpha \rightarrow 1$: fitted exponents go from $\alpha=0.86 \pm 0.02$ for the first incomplete reduction to $\alpha=1.01 \pm 0.01$ for the fully reduced film, as depicted in Fig.\ref{Fig_03}(c, top panel).  We also observed that the value of $\rho_{res}$, usually taken as a signature of the disorder contribution to the resistivity, tends to lower values throughout the reduction, Fig.\ref{Fig_03}(c, bottom panel).  Additional reductions after a fitted exponent $\alpha \approx 1$ has been reached do not increase the critical temperature and might deteriorate the film (see section 7, Fig.S11, Fig.S12 in Supplementary Information).  Moreover, the gradient of linear resistivity in the high-T range upon complete reduction is $A=2.1\pm0.05$ $\mu\Omega\;\mbox{cm}\;\mbox{K}^{-1}$, within the same order of magnitude than $\approx 1.1$ $\mu\Omega\;\mbox{cm}\;\mbox{K}^{-1}$ empirically found across different materials,\cite{bruin:13} including highly crystalline (Nd,Sr)NiO$_2$ films.\cite{lee:23}

Proximity to a quantum critical point, Planckian dissipation and existence of a charge gap (Mott insulation) that remains upon doping have been proposed as underlying principles of linear-in temperature resistivity observed in some metals,\cite{phillips:22} but there is no widely accepted explanation. To further elucidate the resistive behaviour of nickelates superconductors, additional experimental exploration of different manifestations of anomalous behaviour in magnetotransport properties of the normal state, such as $H$-linear magnetoresistance at high field $H$, over a broad doping range, is needed.  Nonetheless, the fact that $T$-linear resistivity is observed in nickelates superconductors that are not doped Mott insulators indicate that Mott insulation of the undoped parent compound is not a necessary ingredient for $T$-linear resistivity. Furthermore, these results support those recently found in nickelates superconductors with enhanced crystallinity on LSAT substrates.\cite{lee:23}

\begin{figure*}[htbp]
 \includegraphics[keepaspectratio=true, width=\linewidth]{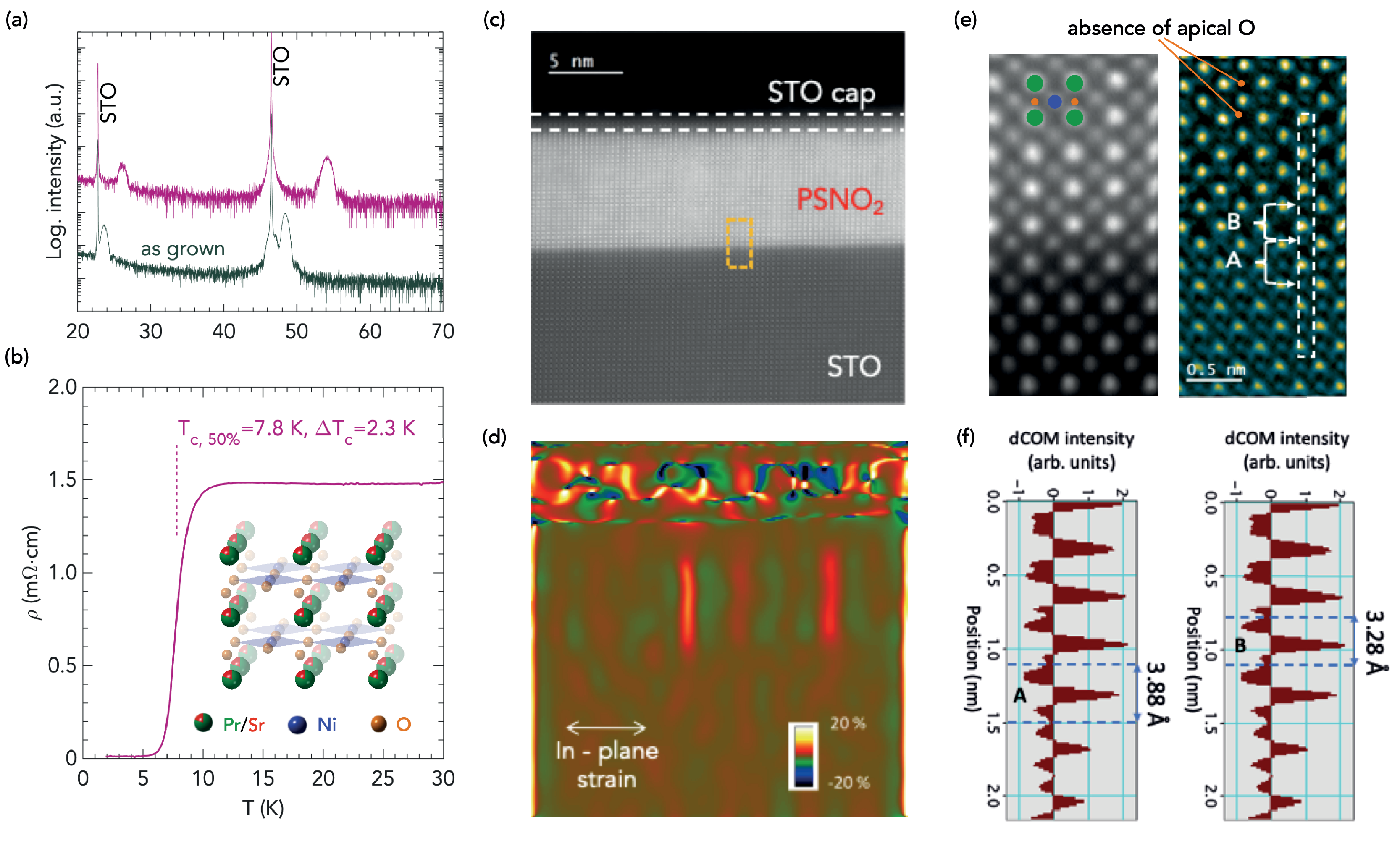} \caption{\label{Fig_04} {{\bf{Structure of SC IL PSNO$_2$ thin film.}}  {\bf{(a)}} XRD $\theta-2\theta$ symmetric scans of a PSNO$_3$ film as-grown (grey) and after reduction (pink), showing reflections consistent with the IL tetragonal PSNO$_2$ phase.  The curves are vertically offset for clarity.  {\bf{(b)}} Temperature dependent resistivity $\rho(T)$ of the reduce sample, showing the SC transition at $T_{c, 50\%}=7.8 \;\mbox{K}$. Inset: schematic of the crystal structure of the IL phase PSNO$_2$ with Ni$^{1.2+}$ in a square-planar environment (tetragonal space group P4/mmm)).   {\bf{(c)}} Atomic-resolution HAADF-STEM image of the IL film, scale bar, 5 nm.  {\bf{(d)}} Geometrical Phase Analysis of the in-plane components of the strain tensor obtained from the STEM image in panel {\textit{c}}. The red regions show possible Ruddlesden–Popper defects.  {\bf{(e)}} Atomic-resolution HAADF-STEM image (left) from the region highlighted in panel {\textit{c}} as a dashed orange rectangle (green circles indicate Pr/Sr atom sites; blue circles, Ni atom sites; and, orange circles, oxygen in the NiO$_2$ planes); and, 4D-STEM dCOM image (right) revealing the absence of apical oxygen anions.  White arrows show positions between which distances are calculated in panel \textit{f}.  {{\bf{(f)}}} Calculated dCOM intensity profile across the PSNO$_2$$|$STO interface along the atomic column highlighted in the 4D-STEM dCOM image (panel \textit{e}, right): oxygen-oxygen distances of 3.88 $\mbox{\AA}$ in STO and 3.28 $\mbox{\AA}$ in PSNO$_2$ are found (locations marked as A and B, respectively, in panel \textit{e}, right). All the measurements were taken from the same sample.  
}}
\end{figure*}

Upon further cooling, we find a resistivity upturn that holds between a temperature $T^{\prime}$, at which $\rho(T)$ reaches its local minimum, and the onset of the SC transition, Fig.\ref{Fig_03}(d).  There is evidence of this behaviour for previously reported SC nickelate thin films on STO substrates.\cite{Li:19,osada:21}  This feature appears to provide additional support for a decrease of disorder in our films throughout the reduction since the temperature $T^{\prime}$ lowers along the reduction process from 34.3 K after the first step to 26.7 K after the third, and reaches 19.8 K after the fourth, narrowing the temperature range over which the resistivity increases. Also, the upturn gradually smooths out and the resistivity curve flattens, nearly vanishing for the fully reduced sample, whose resistivity exhibits a very weak minimum. Additional transport characterization is shown in Supplementary Information (section 8).

Fig.\ref{Fig_03}(e) illustrates the evolution of SC parameters of the reduced sample over successive annealings.  Incremental reduction of the film results in a gradual increase of the onset of superconductivity, $T_{onset}$, and of the critical temperature $T_c$, defined as the temperature at which resistance reaches half the normal state value at 20 K.  The highest $T_{onset}\approx 9\;\mbox{K}$, which corresponds to a critical temperature of $T_{c}=7.6\;\mbox{K}$, and the lowest resistivity $\rho(T_c)=0.2$ $\mbox{m}\Omega\;\mbox{cm}$ (Fig. S8, Supplementary Information) are found after the whole reduction process.  The width of the resistive transition (90\%–10\% of normal state at 20 K) in the absence of an applied magnetic field, $\Delta T_c$, is around 3 K, and slightly decreases throughout the process.  No transition widths are given in literature for SC nickelates, and they are difficult to discern from the plots, but widths of the order of 2 K-3 K are usually found in high quality samples of doped cuprates.\cite{saadaoui:15}

The observed increase in $T_c$ over incremental reduction processes can be attributed to a decrease of disorder, supporting the discussion above. Furthermore, intrinsic inhomogeneity inherent to doping, such as inhomogeneous charge density due to random distribution of dopants and disorder owing to difference in the ionic radius of Pr$^{3+}$ and Sr$^{2+}$, can reduce the attainable $T_c$ as it has been reported in cuprates.\cite{eisaki:04,fujita:05}  Although, on the other hand, nanoscale electronic disorder, studied using scanning tunnelling microscopy and spectroscopy, has been found to coexist with high SC transition temperatures in cuprates.\cite{pan:01,mcElroy:05} In our results, since the dopant distribution is constant along the reduction process, the increase in $T_c$ suggests an overall decrease of structural defects as the reduction progresses.   However, scanning tunnelling microscopy experiments are needed on nickelates to reveal the effect of dopant disorder on their local superconducting properties.

The Residual Resistivity Ratio (RRR), which is a measure of the scattering rate of charge carriers by impurities or defects in the films, and is also thought to be indicative of residual apical oxygen atoms in the NiO$_6$ octahedra in the reduced phase, is depicted in Fig.\ref{Fig_03}(f).  As a reference, the RRR in the as-grown sample was $\approx 11$.  In the reduced films, the RRR increases through the overall reduction process, suggesting a progressive decreasing of structural disorder over the orthorhombic to tetragonal transition along with a gradual removal of apical oxygen anions.  RR ratios of around 2.8 are achieved, approximately equal to those attained elsewhere.\cite{osada:20a,osada:20b}  The significant improvement of the SC transition after the second step does not result in a subsequent increase in the RRR.  This may indicate the formation of a restricted SC region in the sample, surrounded by regions that do not superconduct.   

As shown in Fig.\ref{Fig_03}(g), the normal state sheet resistance measured at 20 K lowers down to $R_s \approx 770 \; \Omega\;\square^{-1}$ ($\approx 0.030$ in $h/e^2$ units, where $h$ is Planck's constant and $e$, elementary charge, taken as an approximative value for the Mott-Ioffe-Regel limit in two dimensions).  The increase of the absolute value $R_s$ after the second reduction may be due to slight uncertainties in the contact sizes in the Van der Pauw measurement and not to a degradation of the film, since the RRR value keeps the same as for the first reduction, and when normalised at 300 K, $\rho/\rho_{300\;K}$, shows a subsequent improvement (Fig.\ref{Fig_03}(a) and Fig. S7,Supplementary Information).  

The removal of oxygen anions in the reduced phase can also be tracked by the shrinkage of the $c$-axis lattice parameter.  Fig.\ref{Fig_03}(h) shows the $c$-axis lattice parameter as function of $T_{onset}$.  We find a striking $\approx$12\% decrease in the lattice parameter along $c$ from the parent perovskite phase to the complete topotactic oxygen deintercalation, and although a precise lattice parameter determination is hindered by the limited number of accessible Bragg reflections, our values are in agreement with those in literature for the SC IL PSNO$_2$ phase.\cite{osada:20a,osada:20b} This contraction along $c$ leads to a decrease in the thickness of the film (estimated from the Scherrer equation), as depicted in Fig.\ref{Fig_03}(i), in agreement with STEM measurements as discussed below.

The structural analysis of the IL phase was completed by STEM on an optimized IL PSNO$_2$ film whose XRD pattern and temperature dependent resistivity are shown in Fig.\ref{Fig_04}(a, b), showing  a critical temperature of $T_{c, 50\%}=7.8\;\mbox{K}$.  Inset in Fig.\ref{Fig_04}(b) depicts the schematic of the crystal structure of the IL phase, where the infinite NiO$_2$ planes are separated only by Pr/Sr atoms, once the apical oxygen anions of the starting perovskite phase have been removed. The cross-sectional STEM image of the IL film in Fig.\ref{Fig_04}(c) shows a high-quality infinite-layer structure.  And, the geometrical phase analysis (GPA) algorithm applied to that STEM image in Fig.\ref{Fig_04}(d) reveals only a few possible Ruddlesden–Popper defects.  Furthermore, 4D-STEM divergence of center of mass (dCOM) image, approximating to a projected charge density image in Fig.\ref{Fig_04}(e) confirms the removal of apical oxygen anions from the perovskite phase so that Pr/Sr planes alternate with NiO$_2$ planes in the film.  We also observed an abrupt oxygen-oxygen distance variation at the interface of the reduced film (Fig.\ref{Fig_04}(f)), changing from $\approx 3.9\;\mbox{\r{A}}$  in the STO substrate (left panel, A)  to $\approx 3.3\;\mbox{\r{A}}$ in the IL PSNO$_2$ film (right panel, B), which confirms the topotactic transformation at the interface.  Moreover, the out-of-plane lattice parameter estimated from the GPA analysis depicted in Fig. S14 corresponds to a decrease of 15\% related to the STO lattice parameter ($3.91\; \mbox{\r{A}}$), closely matching the values determined for the fully reduced sample by X-ray diffraction in Fig.\ref{Fig_03}(i).  The present work opens up the prospect of experimentally studying the structure of the substrate-nickelate interface at atomic scale, which goes beyond the scope of this work. Specifically, questions arise about possible alternative atomic interfaces (see Fig. S15, Supplementary Infor\-ma\-tion), different to that previously reported,\cite{goodge:23} and whether the topotactic transformation might result in interface reconstructions revealed by first principles calculations for Nd-nickelate\cite{geisler:20} or La-nickelate.\cite{bernardini:20}

We finally mention that attempts to obtain SC films from uncut samples were always unproductive and a deeper understanding of the topotactic process is needed to explain this empirical observation. Immediately following unsealing of the ampoule, the uncut film shows XRD patterns or temperature dependence of resistivity typical of a reduced phase, but it readily reoxidizes in less than 24 hours even when stored in a glovebox under nitrogen atmosphere or under vacuum (see Supplementary Information, section 11, Fig. S16 to Fig. S18).

\section{\label{sec:conclusions}Conclusions}

In summary, we have accomplished the synthesis of SC infinite-layer IL praseodymium nickelate thin films.  Our results highlight the importance of the combined optimization of both steps of the process: perovskite growth, with control over cation stoichiometry, and topochemical reduction.  Starting from nearly stoichiometric perovskite films with minimal defect densities, the linear-in-temperature resistivity of intermediate reduced films can be used as a proxy to assess the performance of subsequent reduction processes.  These results contribute towards the goal of yielding high quality superconducting nickelate samples, still scarce to date, to enable further experimental progress in this field.  Furthermore, better understanding of the topochemical aspects of the reduction process is a critical issue for exploring this new class of superconductors and could push forward experimental research on the field.

\section{\label{sec:Experimental}Experimental methods}

\subsection*{Pulsed Laser Deposition and CaH\texorpdfstring{\textsubscript 2}{2} reduction}

We grew perovskite nickelate thin films in a PLD system which utilizes a KrF excimer laser (248 nm) focused onto the target. The laser pulse rate was fixed at 4 Hz. During the growth, oxygen was supplied in the PLD chamber yielding a background pressure up to 0.5 mbar. The laser fluence was varied between 1.2 $\mbox{J}\,\mbox{cm}^{-2}$ and 2.5 $\mbox{J}\,\mbox{cm}^{-2}$, with a laser spot size of $2\pm 0.2 \;\mbox{mm}^2$.  The substrate temperature was set in the range 570 $^{\circ}$C - 675 $^{\circ}$C. We optimized these parameters to produce high-quality epitaxial films as discussed in the text. The films were cooled down to room temperature at a rate of 5 $^{\circ}\mbox{C}\,\mbox{min}^{-1}$ at the growth pressure. Single crystals of (001)SrTiO$_3$ (STO) were used as substrates. Prior to the growth, they were etched in buffered HF and annealed at 1000 $^{\circ}$C for 3 h to obtain stepped surfaces. We sintered the PLD target from a mixture of Pr$_2$O$_3$ (99.99\%), Nickel(II) oxide (99.99\%) and  SrCO$_3$ with controlled cation stoichiometry ([Ni]/([Pr]+[Sr])=1.1; [Pr]/[Sr]=4) by a solid state reaction.  These mixtures were ground in an agate mortar and, after initial decarbonation at 1200 $^{\circ}$C for 12 h, pressed into pellets, and heated in a box furnace at 1300 $^{\circ}$C for 24 h. To get high-density PLD targets, the powders were reground and repressed, and then fired at 1300 $^{\circ}$C for further 24 h.

Reduction of the perovskite phase into the IL phase was carried out in evacuated glass tubes. The tubes were filled with 0.1 g of CaH$_2$ powder in an N$_2$-filled glovebox. Samples were wrapped in aluminum foil and inserted into the glass tubes.  The samples are separated from the CaH$_2$ powder (Alfa Aesar A16242) by means of a lump of glass wool (see Fig. S19, Supplementary Information). The tubes were evacuated by means of a rotary pump ($< 10^{-3}$ mbar) and sealed.  A horizontal tube furnace,  ventilated with N$_2$ for increased temperature homogeneity was used for the process, with temperature accuracy $\pm 1\; ^{\circ}$C.  The reference temperature is measured at the center of the tube furnace, where the sealed ampoule is placed.  Heating and cooling rates in the oven were 5 $^{\circ}\mbox{C}\;\mbox{min}^{-1}$.

\subsection*{X-ray Diffraction}

XRD $\theta$/2$\theta$ scans of the films were performed in a four-circle diffractometer (Cu source, Ge(220) 2-bounce incident beam monochromator).  The lattice parameters are calculated from Gaussian fits of the (001) and (002) XRD peaks (indices with respect to the pseudocubic unit cell) and extrapolated against $\cos^2\theta/\sin \theta$ to $\theta=90^{\circ}$ to reduce systematic errors.  The thickness of the reduced films was estimated from the width of the (002) XRD reflection through the Scherrer equation with a constant $K=1.06$ fitted for our samples.  Scans around asymmetrical reflections of the films were transformed to Reciprocal Space Maps (RSMs).

\subsection*{X-ray Photoelectron Spectroscopy}

XPS measurements (\textit{in situ} and \textit{ex situ}) were performed using a Mg K$\alpha$ source (1253.6 eV, 20 mA, 15 kV).  Survey spectra were acquired with a pass energy of 60 eV and detailed spectra with a pass energy of 20 eV in the energy analyzer. All spectra were measured in normal emission. XPS data were processed with the CasaXPS software. 

\subsection*{Electrical transport}

Transport measurements were performed in a Physical Property Measurement System (PPMS, Quantum Design).  Four-point resistivity measurements were performed in a Van der Pauw geometry by means of wire-bonded Au wires.  Temperature-dependent Hall coefficients were calculated from linear fits of antisymmetrized field sweeps up to 9 T.

\subsection*{Transmission Electron Microscopy}

The cross-sectional lamellae for Transmission Electron Microscopy were prepared using a Focused Ion Beam (FIB) technique at Centre de Nanosciences et de Nanotechnologies (C2N), University Paris-Saclay, France. Prior to FIB lamellae preparation, around 20-30 nm of amorphous carbon was deposited on top of the samples for protection.  The High-angle annular dark-field (HAADF) imaging and 4D-STEM was carried out in a NION UltraSTEM 200 C3/C5-corrected scanning transmission electron microscope (STEM).
The experiments were done at 200 keV with a probe current of $\approx$12 pA and convergence semi-angles of 23 mrad. A MerlinEM (Quantum Detectors Ltd) in a 4 × 1 configuration (1024 × 256) had been installed on a Gatan ENFINA spectrometer mounted on the microscope.\cite{tence:20} For 4D-STEM, the electron energy loss spectroscopy (EELS) spectrometer was set into non-energy dispersive trajectories and 6-bit detector mode that gave a diffraction pattern with a good signal to noise ratio without compromising much on the scanning speed was used. The geometrical phase analysis (GPA)\cite{hytch:98} had been done choosing the STO substrate with $3.91 \;\mbox{\AA}$ as a reference parameter. The lattice parameters of the PSNO$_2$ were estimated by averaging the GPA maps over square areas of $\approx$ 50 (in-plane) $\times$ 50 (out-of-plane) nm giving a strain accuracy determination better than 1\%, that is, better than $0.04 \;\mbox{\AA}$ for the lattice parameters.  Such an approach has been previously employed to accurately determine the $c$-axis variation in an apical oxygen ordered nickelate thin-film on an STO substrate.\cite{raji:23}  The EELS spectra were obtained using the full 4 $\times$ 1 configuration and the 4D-STEM by selecting only one of the chips (256 $\times$ 256 pixels). The element maps were done by integrating the core EELS edge-signal of the respective elements and mapping them in the spectrum image.

\subsection*{Acknowledgments}

The authors thank Jin-Hong Lee for his help at the early stage of this project and Richard Lebourgeois for his support with target preparation.  A.G.L. acknowledges financial support through a research grant from the Next Generation EU plan 2021, European Union.  D. Z. acknowledges finantial support from \'{E}cole Doctorale 564 Physique en Ile de France (EDPIF).  This work received funding from the ERC Advanced Grant No. 833973 (Fresco).

\subsection*{Additional Information}

Supporting Information is available for this manuscript.

\subsection*{Author contributions}

A.G.L. designed and performed the experiments (growth, XRD, XPS, topochemical reductions, electrical transport), processed and analysed the data, produced the graphics and wrote the manuscript.  A.R. and A.G. designed and performed electron microscopy experiments, processed and analysed the data.  D.Z. and F.G. performed complementary experiments.  L.D. and C.G. designed and implemented the experimental setup for sealing the glass ampoules used in the reduction process.  L.I. initiated the optimization (growth, XRD, topochemical reduction, electrical transport) of nickelate samples at the beginning of the project and assisted in reduction experiments.  M.B. proposed the project, provided equipement and acquired funding.  A.G.L., A.R., D.Z., A.G., L.I. and M.B. discussed the data.  All co-authors reviewed the manuscript, and approved its final form.

\subsection*{Competing interests}

The authors declare no competing interests.

\subsection*{Data availability}

The data that support the findings of this study are available from the corresponding author upon reasonable request.

\subsection*{Keywords}

nickelates, superconductivity, thin films, topochemical, reduction

\bibliography{ms_2401_12307.bib}

\end{document}

\title{{\small{Supplemental Material}}\\Towards reliable synthesis of superconducting infinite layer nickelate thin films by topochemical reduction}

\author{Araceli \surname{Guti\'{e}rrez--Llorente}}
\email[]{araceli.gutierrez@urjc.es}
%%\altaffiliation[]{xxx}
\affiliation{Universidad Rey Juan Carlos, Escuela Superior de Ciencias Experimentales y Tecnolog\'{i}a, Madrid 28933, Spain}
\affiliation{Laboratoire Albert Fert, CNRS, Thales, Universit\'{e} Paris Saclay, 91767 Palaiseau, France}

\author{Aravind Raji}
\affiliation{Universit\'{e} Paris Saclay, CNRS, Laboratoire de Physique des Solides, 91405 Orsay, France}
\affiliation{Synchrotron SOLEIL, L’Orme des Merisiers, BP 48 St Aubin, Gif sur Yvette, 91192, France}

\author{Dongxin Zhang}
\affiliation{Laboratoire Albert Fert, CNRS, Thales, Universit\'{e} Paris Saclay, 91767 Palaiseau, France}

\author{Laurent Divay}
\affiliation{Thales Research \& Technology France, 91767 Palaiseau, France}

\author{Alexandre Gloter}
\affiliation{Universit\'{e} Paris Saclay, CNRS, Laboratoire de Physique des Solides, 91405 Orsay, France}

\author{Fernando Gallego}
\affiliation{Laboratoire Albert Fert, CNRS, Thales, Universit\'{e} Paris Saclay, 91767 Palaiseau, France}

\author{Christophe Galindo}
\affiliation{Thales Research \& Technology France, 91767 Palaiseau, France}

\author{Manuel Bibes}
\affiliation{Laboratoire Albert Fert, CNRS, Thales, Universit\'{e} Paris Saclay, 91767 Palaiseau, France}

\author{Luc\'{i}a Iglesias}
\email[]{lucia.iglesias@cnrs-thales.fr}
\affiliation{Laboratoire Albert Fert, CNRS, Thales, Universit\'{e} Paris Saclay, 91767 Palaiseau, France}

%\date{\today}

\maketitle %\maketitle must follow title, authors, abstract and \pacs

\tableofcontents

%\section{\label{sec:intro}xxxxx}

\subsection{Optimal growth of PSNO\texorpdfstring{\textsubscript 3}{3} films}

\begin{figure*}[h]
 \includegraphics[keepaspectratio=true, width=\linewidth]{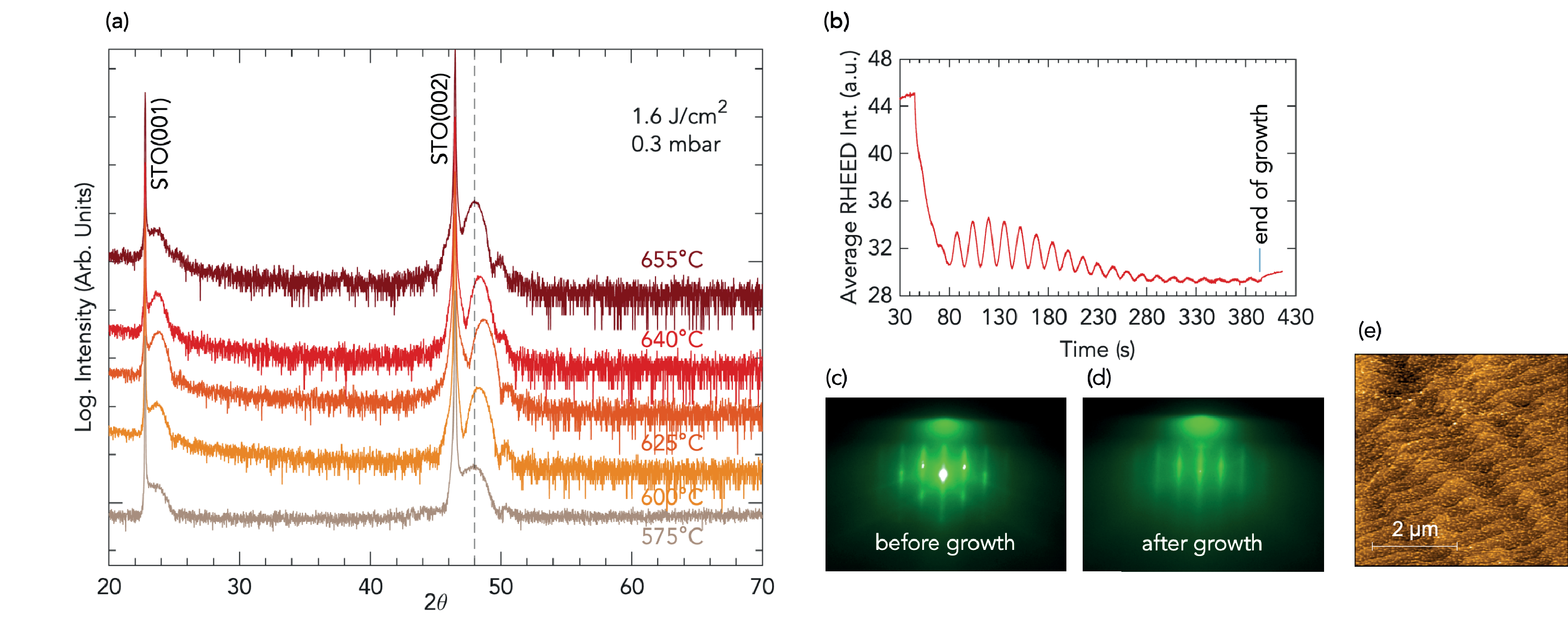} \caption{\label{Fig_S1_01} {{\bf{(a)}} XRD symmetric $\theta-2\theta$ scan patterns of PSNO$_3$ films grown on STO substrates at the optimized conditions of laser fluence 1.6 J/cm$^2$, oxygen pressure 0.3 mbar, and substrate temperatures ranging from 575 $^{\circ}$C to 655$^{\circ}$C.  The dashed horizontal line is an estimate of $c$ from the bulk lattice constant of PNO$_3$ strained to the STO substrate, and doping with Sr is expected to bring about a contraction of the unit cell.  {\bf{(b)}} RHEED intensity oscillations observed during the growth of a PSNO$_3$ film under optimal conditions of 1.6 J/cm$^2$, 640$^{\circ}$C, 0.3 mbar. RHEED diffraction patterns taken {\bf{(c)}} before, and {\bf{(d)}} after the growth shown in panel (b). {\bf{(e)}} AFM image of the as-grown film. 
}}
\end{figure*}

%(indices with respect to the pseudocubic unit cell)

\clearpage

\subsection{Low-temperature resistivity of PSNO\texorpdfstring{\textsubscript 3}{3} films grown under non-optimal conditions}

Structural or composition disorder results in a decrease of the mean free path of the carriers, and quantum mechanical corrections to the low-temperature conductivity should be taken into account, since even in the weak-disorder limit the Boltzmann description $\rho(T)=\rho_0(T)$ is no longer valid.

We have fitted the experimental resistivity data at low-temperature to the temperature dependence\cite{herranz:05} 

\begin{equation}\label{eq_s01}
\rho(T)=\frac{1}{\sigma_0 + a \ln(T)}+b T^2
\end{equation}

where $\sigma_0$ is the residual conductivity due to scattering by defects, and is a measure of the degree of disorder; the term propotional to $\sim \ln$ gives the quantum correction of the conductivity; and, the term $b T^2$ accounts for the classical low-temperature dependence of the resistivity. Table \ref{table_S01} shows the fitting parameters to that behaviour.  We observe higher degree of disorder in the film grown at higher laser fluence, with the minima in resistivity occurring at a highest temperature.

\begin{figure*}[h]
 \includegraphics[keepaspectratio=true, width=\linewidth]{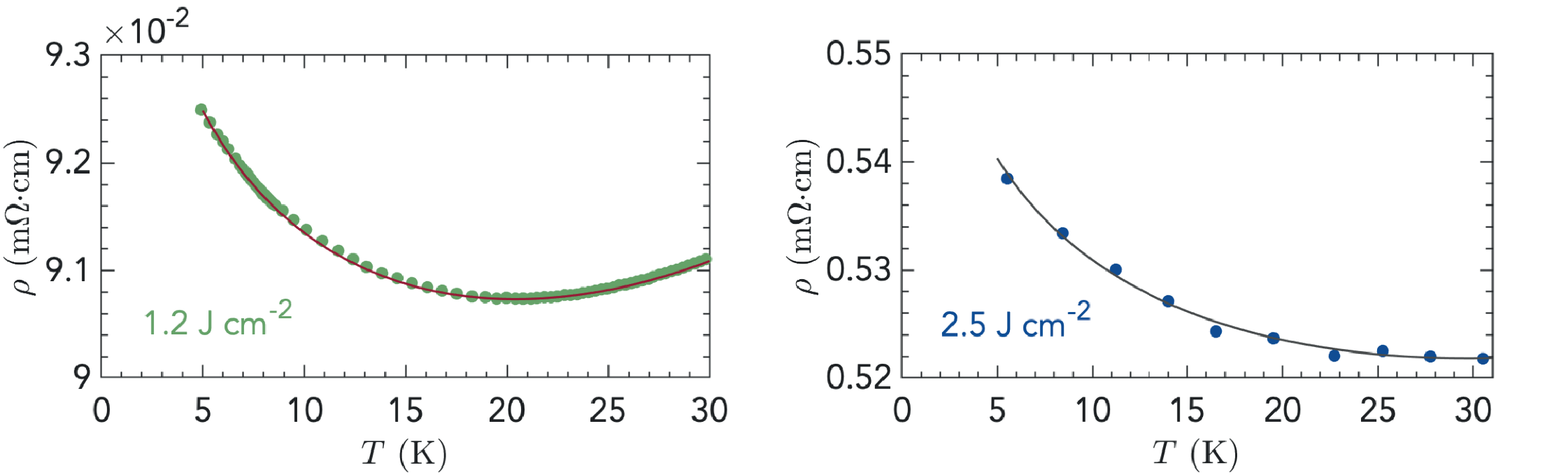} \caption{\label{Fig_S0200_rev} {Temperature dependence of resistivity,$\rho(T)$, at low temperatures in PSNO$_3$ films grown at 600 $^{\circ}$C, 0.4 mbar and a laser fluence of 1.2 $\mbox{J}\,\mbox{cm}^{-2}$ (left panel) and 2.5 $\mbox{J}\,\mbox{cm}^{-2}$ (right panel). Solid circles are experimental data and solid lines are least-square fits of the experimental data to Eq.\ref{eq_s01}.   
}}
\end{figure*}

\begin{table}[htb]
\caption{Fitting parameters of experimental resistivity as a function of temperature to the Eq.\ref{eq_s01} in the temperature range from 5 K to 30 K. Fits are shown in Fig.\ref{Fig_S0200_rev}}.
\centering
\setlength{\tabcolsep}{10pt}
\renewcommand{\arraystretch}{1.5}
\begin{tabular}{ccccc}
%\hline \hline
laser fluence & $T_{min}$ (K) & $  \sigma_0$ & $a$ & $b$  \\
\hline %\hline
%&  &  &  &  \\
1.2 $\mbox{J}\,\mbox{cm}^{-2}$ & 10.8 & 10.47 & 0.22 & $2.3 \cdot 10^{-6}$ \\ 
2.5 $\mbox{J}\,\mbox{cm}^{-2}$ & 30.5 & 1.77 & 0.05 & $8.3 \cdot 10^{-6}$ \\
\hline
\end{tabular}
\label{table_S01}
\end{table}

These quantum-mechanical corrections have two contributions: localization of the wave function and electron-electron interaction, leading both of them to an increase of the resistivity as the temperature decreases with very similar dependence on temperature (either in 3D or 2D models).\cite{lee:85}  Differentiating both contributions would require careful analysis of the field dependence of the low-temperature resistivity,\cite{herranz:05} which is beyond the scope of this paper.

\clearpage

\subsection{Cation stoichiometry of the PSNO\texorpdfstring{\textsubscript 3}{3} films by X-ray Photoelectron Spectroscopy}

\begin{figure*}[!htb]
 \includegraphics[keepaspectratio=true, width=0.85\linewidth]{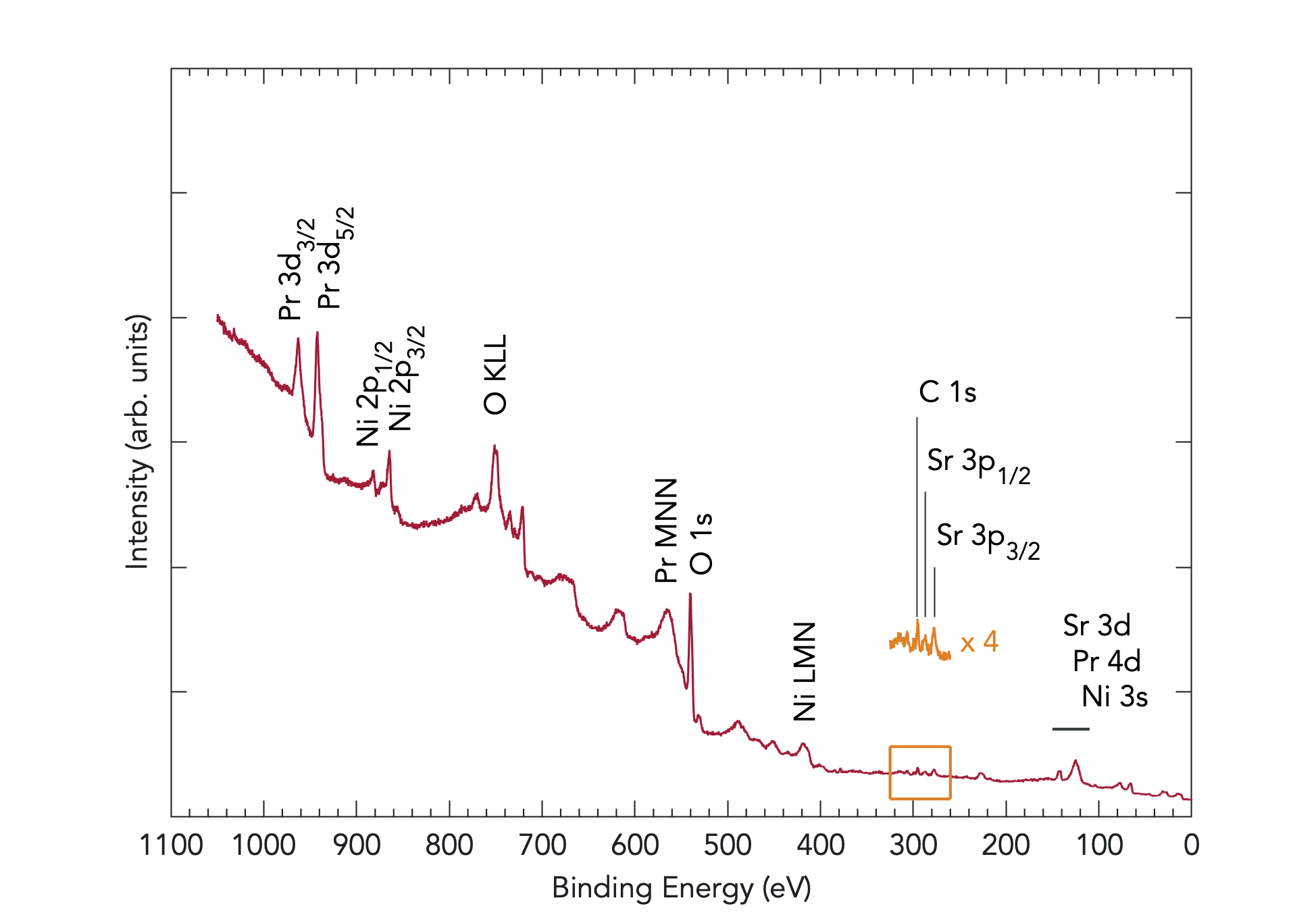} \caption{\label{Fig_XPS_survey} {Mg K$\alpha$-excited XPS survey spectrum performed at high pass energy (60 eV) on a perovskite SPNO$_3$ film showing the assignments of the peaks.  Pr 3d and Ni 2p core-levels are used for quantification of the $\mbox{[Pr]/[Ni]}$ ratio, as detailed below.  Quantification of Sr is done from Sr 3p$_{3/2}$ since Sr 3p$_{1/2}$ overlaps the C 1s region, and Sr 3d core level overlaps Pr 4d and Ni 3s regions.
}}
\end{figure*}

\begin{figure*}[!htb]
 \includegraphics[keepaspectratio=true, width=0.8\linewidth]{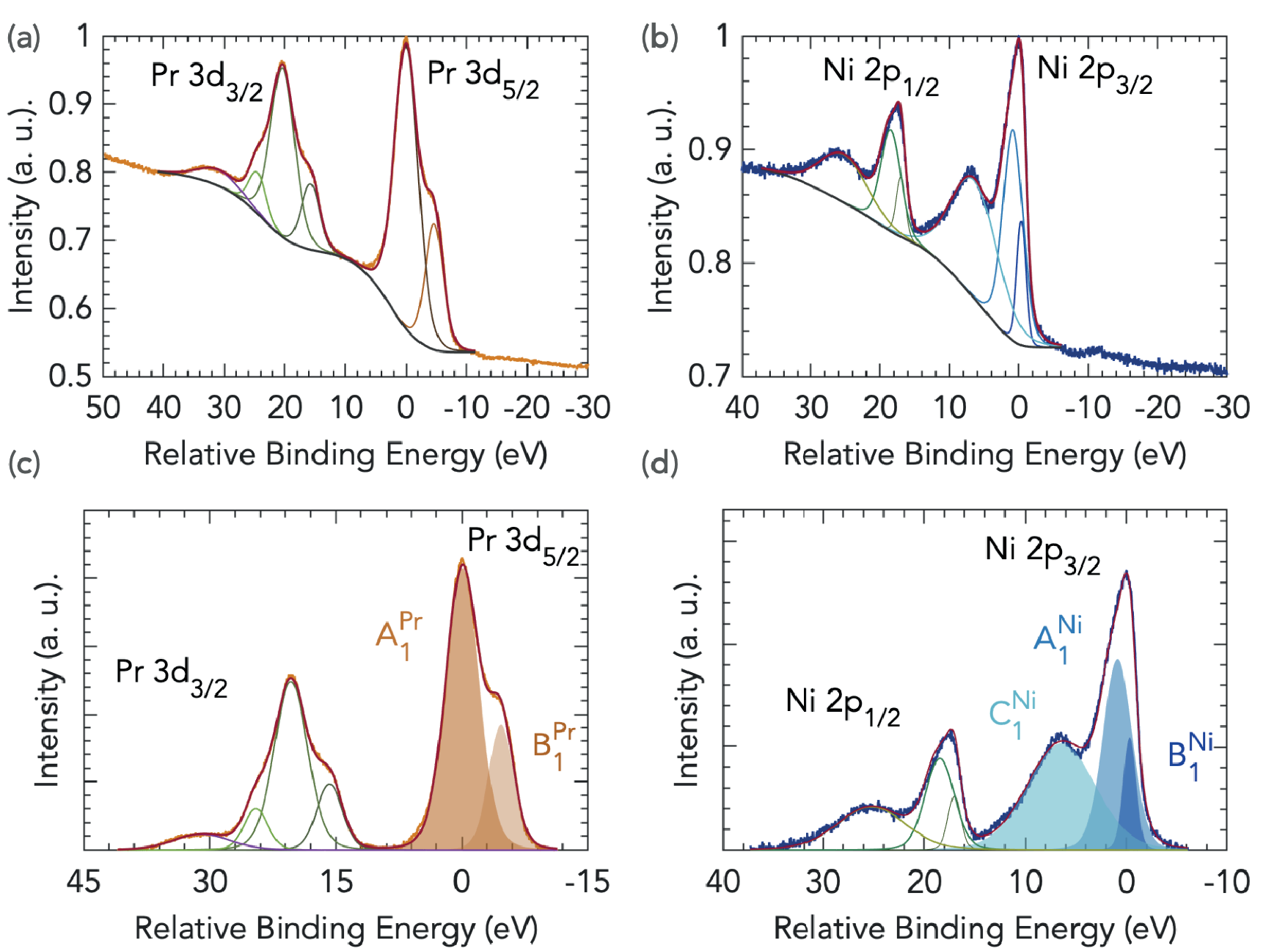} \caption{\label{Fig_SI_XPS}{{\bf{(a, b)}} Examples of Pr 3d and Ni 2p doublets core-level XPS spectra, respectively, from a PSNO$_3$ film fitted with a Tougaard background extending over both components of each doublet. {\bf{(c, d)}} Peak models used for the quantification of the $\mbox{[Pr]/[Ni]}$ ratio. The components ((c) brown areas, Pr; (d) blue areas, Ni) and total fit envelope are shown. 
}}
\end{figure*}

Quantitative XPS for the determination of the $\mbox{[Pr]/[Ni]}$ ratio was derived from the area under the core-level and satellite peaks, Pr 3\textit{d} and Ni 2\textit{p}.  The Pr 3\textit{d} core level spectra consist of a spin-orbit split doublet with Pr 3$d_{5/2}$ and Pr 3$d_{3/2}$ components, at 3:2 area ratio, and Ni 2\textit{p} core level spectra consist of a doublet Ni 2$p_{3/2}$ and Ni 2$p_{1/2}$, at 2:1 ratio, resolved in energy as illustrated in Fig.\ref{Fig_SI_XPS}(a) and (b), respectively.  Peak areas were obtained after subtraction of a Tougaard background extending over both components for each doublet.\cite{tougaard:21}  To minimize impact of sample charging observed during data acquisition, samples were surrounded with conductive silver paste to provide surface conductive paths during measurements and ensure the best peak shape.  Yet spectra are shown as function of relative binding energy (BE) having the position at the maximum envelope as the zero of the energy scale, since relative peak positions were observed to be stable, and we were not concerned with absolute values of BE but with quantitation using relative intensities.

Fig. \ref{Fig_SI_XPS}(c, d) display peak models used for quantitative analysis of the $\mbox{[Pr]/[Ni]}$ ratio through deconvolution into several component peaks.  To obtain reasonably accurate fitting with meaningful results, the number of component peaks was kept to the minimum that enables an appropriate fit, and constraints were imposed across core-level spectra.\cite{shard:20,brundle:20}  The spin-orbit splitting energies for Pr 3$d_{5/2}$-Pr 3$d_{3/2}$ and Ni 2$p_{3/2}$-Ni 2$p_{1/2}$ were $\Delta E_{\mbox{so}}(Pr)=20.5\; \mbox{eV}\pm0.1\;\mbox{eV}$ and $\Delta E_{\mbox{so}}(Ni)=17.3\; \mbox{eV}\pm0.1\;\mbox{eV}$, respectively, in agreement with literature;\cite{ogasawara:91,fu:18} peak widths were constrained to have the same values in both components of a doublet within 10\%, and peak area ratios between spin-split components were constrained to their theoretical ratio.

The Pr 3$d_{5/2}$ spectrum was fitted with two components: the main peak, $A_1^{Pr}$, that can be attributed to Pr$^{3+}-4f^2$ final state, and its low binding energy satellite peak, $B_1^{Pr}$, Fig.\ref{Fig_SI_XPS}(c).  The separation between those components was kept at $4.6\; \mbox{eV}\pm0.1\;\mbox{eV}$.  The Pr 3$d_{3/2}$ spectrum is more complicated.  Besides the doublets related to the main{} component and its satellite, $A_2^{Pr}$ and $B_2^{Pr}$, an additional extra structure appears at higher BE, that exists only in Pr 3$d_{3/2}$ (not in Pr 3$d_{5/2}$) and has been assigned to a multiplet coupling effect.\cite{ogasawara:91,dash:18}  Only Pr 3$d_{5/2}$ spectrum was used for quantitative analysis in this work.  No evidence of $\mbox{Pr}^{4+}\;(4f^1)$ was found in our XPS spectra, discerned in other Praseodymium compounds as a spectral feature between the Pr 3$d_{5/2}$ and Pr 3$d_{3/2}$ regions,\cite{yaremchenko:14,Gurgul:13,dash:18} where our samples show no signal above the background.  This is in agreement with a recent work on the role of $\mbox{Pr}\;4f$ orbitals on the electronic structures of the undoped and Sr-doped PrNiO$_2$ by density functional theory calculations,\cite{liao:23} where no sign of mixed valency for Pr was found and Pr 4$f$ states were insulating without any hybridization channels near the Fermi energy.  The Ni 2$p_{3/2}$ spectrum was fitted with three components, $A_1^{Ni}$ and $B_1^{Ni}$, whose separation was kept at $1.3\; \mbox{eV}\pm0.1\;\mbox{eV}$, and a broad peak at higher BE, labelled $C_1^{Ni}$ in Fig.\ref{Fig_SI_XPS}(d).

Photoionization cross sections calculated by Scofield\cite{scofield:73} lead to sensitivity factors of 30.72 for $\mbox{Pr} (3d_{5/2})$ and 13.92 for $\mbox{Ni} (2p_{3/2})$ that have been applied in Fig. 2 of the main text. 

%Moreover, an oxygen Auger peak has been reported to appear at the high-energy BE side of Pr 3$d_{3/2}$ as well, and may have an influence in the area below the Pr 3$d_{3/2}$ spectrum.  Thus,

\clearpage

\begin{figure*}[hb]
 \includegraphics[keepaspectratio=true, width=\linewidth]{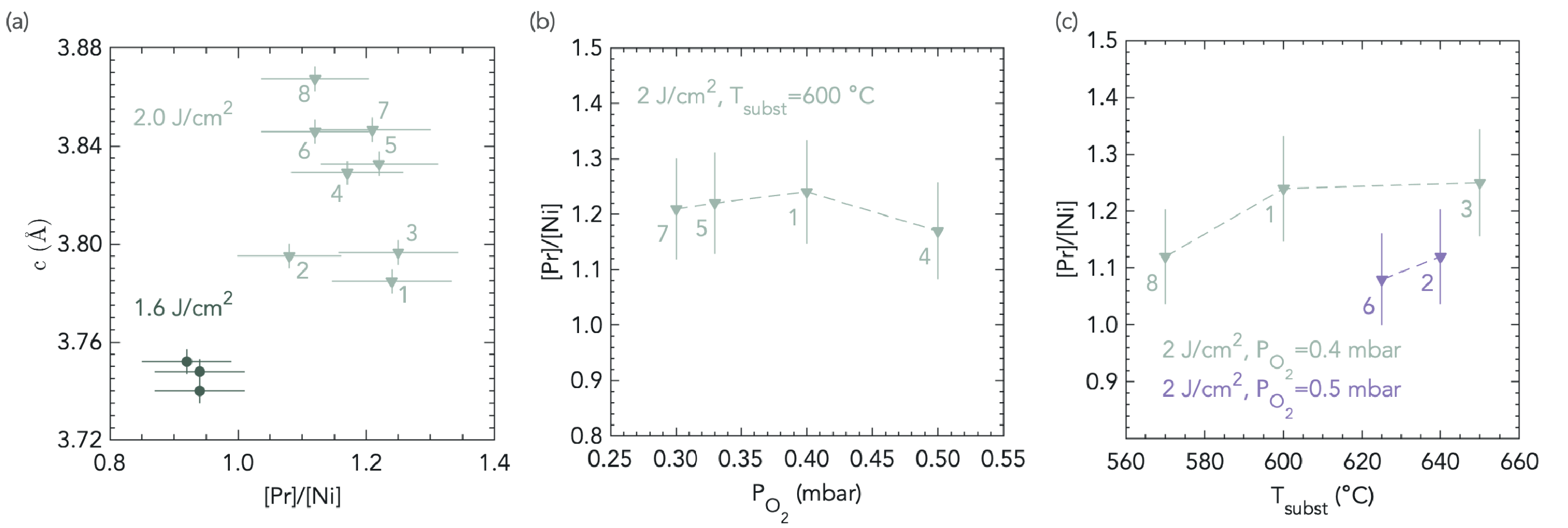} \caption{\label{Fig_Pr_Ni_ratio_at_40mJ} {Exploration of cation stoichiometry of films grown at high laser fluence (2 Jcm$^{-2}$).  {\bf{(a)}} $c$ lattice parameter \textit{vs} $\mbox{[Pr]/[Ni]}$ ratio of PSNO$_3$ films grown at laser fluence 1.6 $\mbox{J}\,\mbox{cm}^{-2}$, 0.3 mbar and substrate temperatures of 625 $^{\circ}$C or 640 $^{\circ}$C, or under non optimal growth conditions of laser fluence 2 $\mbox{J}\,\mbox{cm}^{-2}$ for different series of films, extracted from the fitted XPS spectra.  {\bf{(b,c)}}. $\mbox{[Pr]/[Ni]}$ ratio of films as a function of oxygen pressure (b) and substrate temperature (c) for the series of films shown in (a).  Error bars of 15\% are applied to the $\mbox{[Pr]/[Ni}$ ratio, which is the expected accuracy of XPS quantification for transition metal oxides.\cite{brundle:20}.   
}}
\end{figure*}

\begin{figure*}[!htb]
 \includegraphics[keepaspectratio=true, width=\linewidth]{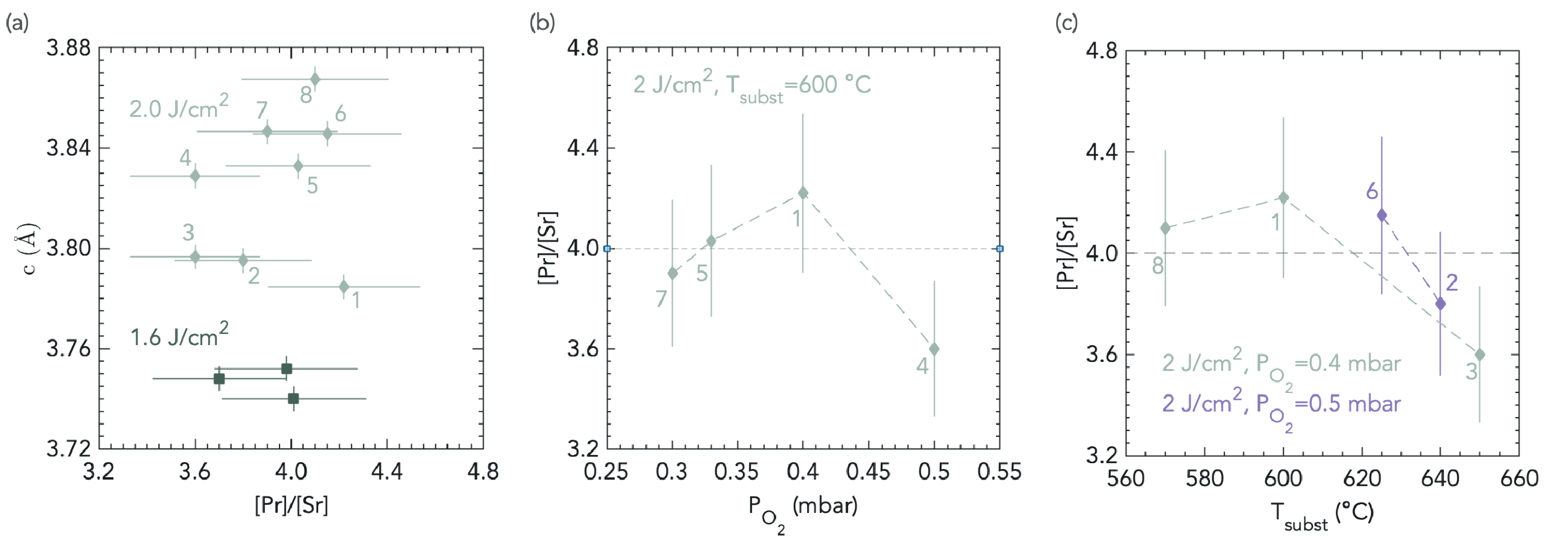} \caption{\label{Fig_XPS_ratios}  {Exploration of cation stoichiometry of films grown at high laser fluence (2 Jcm$^{-2}$). {\bf{(a)}} $c$ lattice parameter \textit{vs} $\mbox{[Pr]/[Sr]}$ ratio of PSNO$_3$ films grown at laser fluence 1.6 $\mbox{J}\,\mbox{cm}^{-2}$, 0.3 mbar and substrate temperatures of 625 $^{\circ}$C or 640 $^{\circ}$C, or under non optimal growth conditions of laser fluence 2 $\mbox{J}\,\mbox{cm}^{-2}$ for different series of films, extracted from the fitted XPS spectra.  {\bf{(b,c)}}. $\mbox{[Pr]/[Sr]}$ ratio of films as a function of oxygen pressure (b) and substrate temperature (c) for the series of films shown in (a).  Error bars of 15\% are applied to the $\mbox{[Pr]/[Sr]}$ ratio in all the panels, which is the expected accuracy of XPS quantification for transition metal oxides.\cite{brundle:20}
}}
\end{figure*}
%grown at pressure: from 0.3 mbar to 0.5 mbar at a temperature of 600 $^{\circ}$C, detailed in panel (b), and temperature between 570 $^{\circ}$C and 650 $^{\circ}$C at 0.4 mbar or 0.5 mbar, detailed in panel (c) }
%\clearpage

%\subsection{Growth of PSNO\texorpdfstring{\textsubscript 3}{3} films: Exploration of growth conditions at high laser fluence}

%\clearpage

%\subsection{Complementary EELS maps of PSNO\texorpdfstring{\textsubscript 3}{3} films}

%\begin{figure*}[h]
 %\includegraphics[keepaspectratio=true, width=0.4\linewidth]{Fig_SI_EELS_maps_perovskite_03} \caption{\label{Fig_EELS_mapping_Sr} {Atomic-resolution HAADF-STEM image and simultaneously recorded elemental EELS maps of Ti L$_{2, 3}$, Pr M$_{4, 5}$ and Sr L$_{2, 3}$ edges of a perovskite PSNO$_3$ film.
%}}
%\end{figure*}

\clearpage

%\clearpage

\subsection{Resistivity of SC PSNO\texorpdfstring{\textsubscript 2}{2} films}

\begin{figure*}[htb]
 \includegraphics[keepaspectratio=true, width=0.55\linewidth]{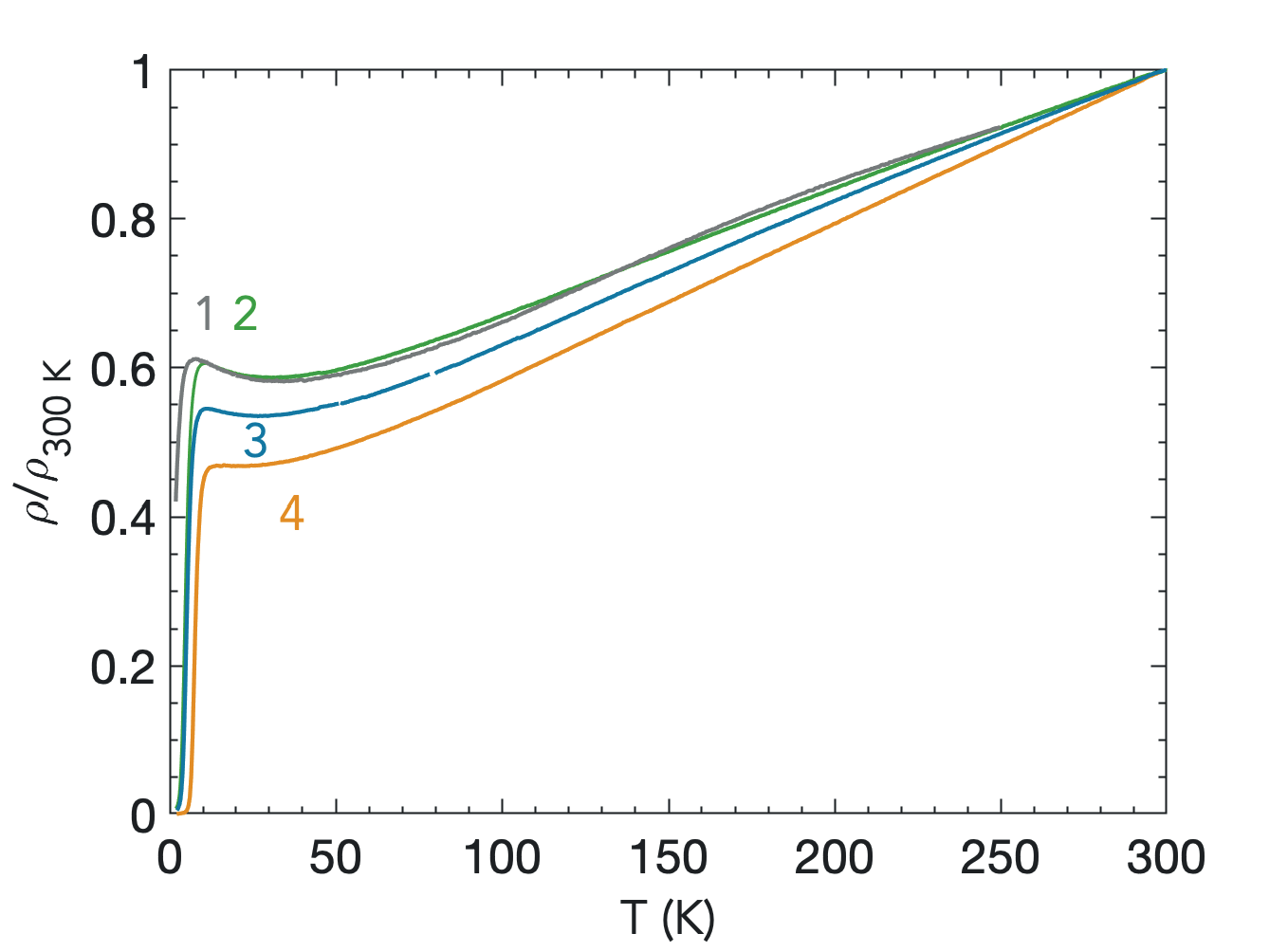} \caption{\label{Fig_rho_norm_300K} {Normalised resistivity in the range from 0 K to 300 K of samples analysed in Fig. 4 of the main manuscript.  Color code as in the main figure.  
}}
\end{figure*}

\begin{figure*}[h]
 \includegraphics[keepaspectratio=true, width=0.55\linewidth]{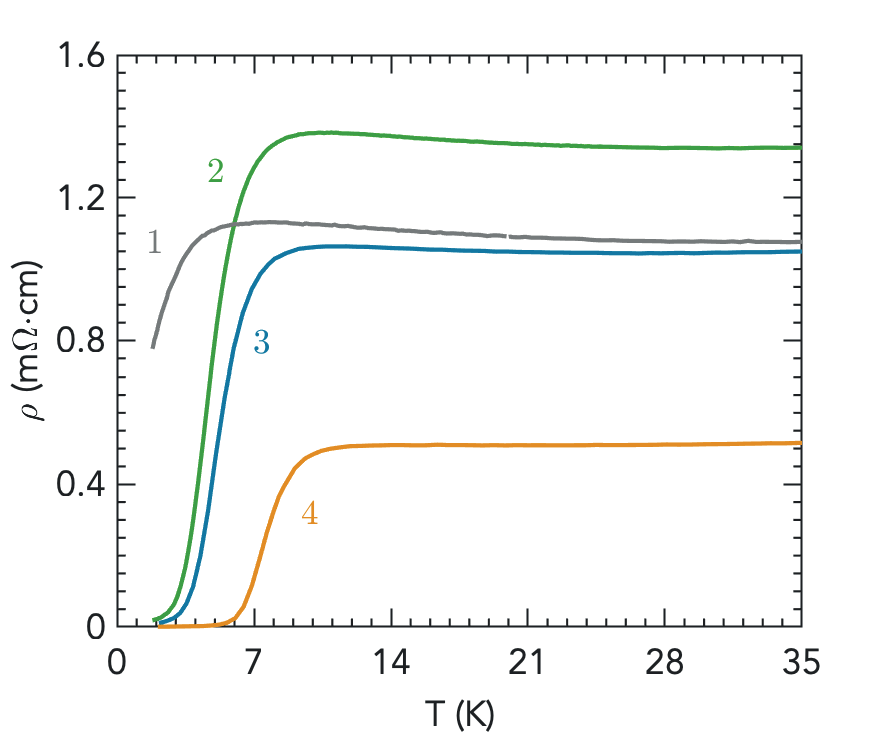} \caption{\label{Fig_raw_rho} {Unnormalized resistivity of samples analysed in Fig. 4 of the main manuscript.  Color code as in the main figure.  
}}
\end{figure*}

\clearpage

%\subsection{XRD patterns of SC PSNO\texorpdfstring{\textsubscript 2}{2} films}

\subsection{XRD patterns of topochemically reduced films}

\begin{figure*}[htb]
 \includegraphics[keepaspectratio=true, width=0.6\linewidth]{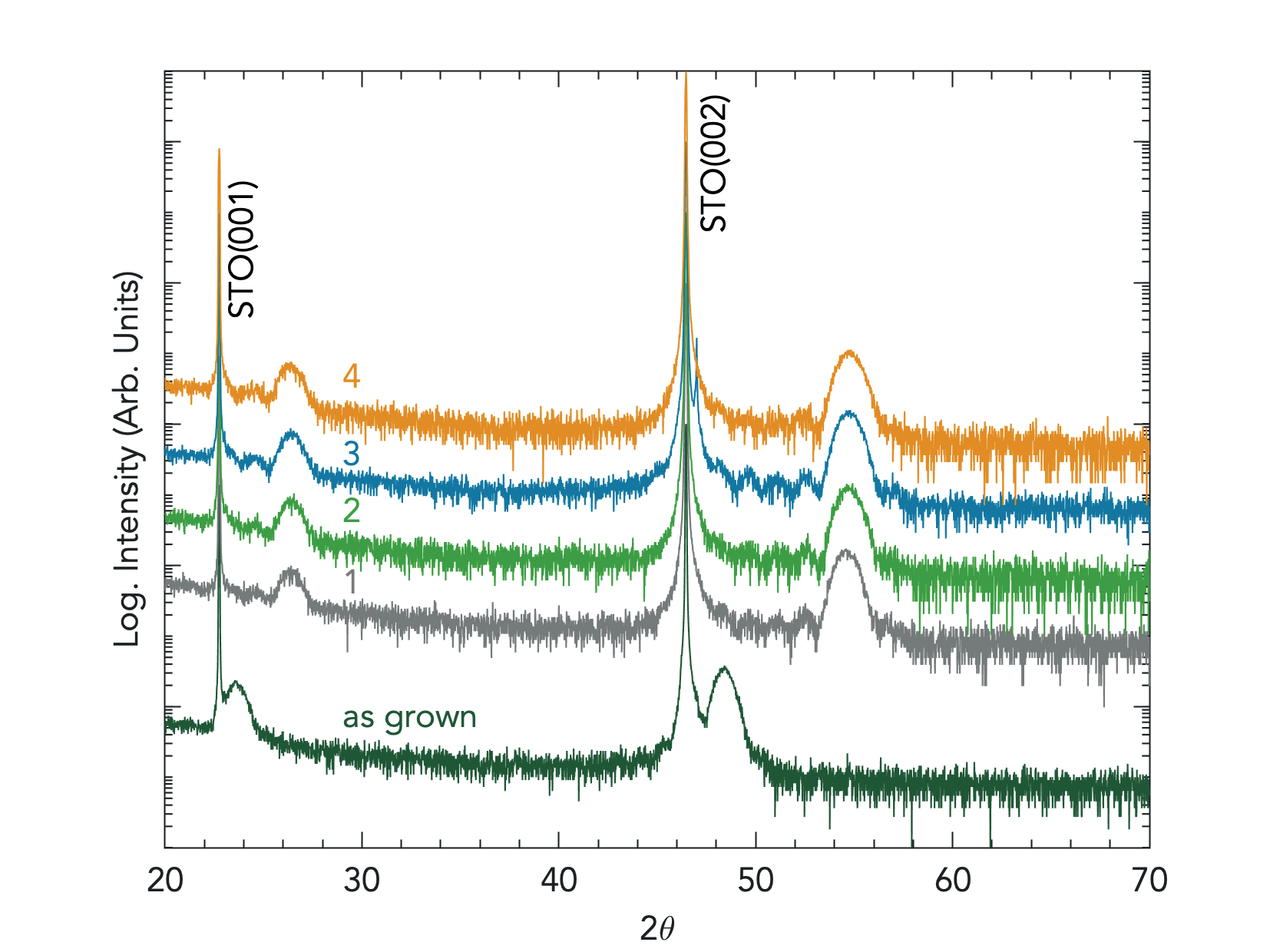} \caption{\label{Fig_xrd_reductions} {XRD $\theta-2\theta$ symmetric scans of the as-grown and reduced films after consecutive reduction processes carried out at 260$^{\circ}$C for periods of (1) 150 min, and additional 30 min (2), 25 min (3), and 45 min (4), whose superconducting parameters are shown in Fig. 4 of the main text. The curves are vertically offset for clarity.  Color code as in the main figure.
}}
\end{figure*}

%\clearpage

\subsection{Fits of the normal-state resistivity of SC PSNO\texorpdfstring{\textsubscript 2}{2} films}

\begin{figure*}[h]
 \includegraphics[keepaspectratio=true, width=\linewidth]{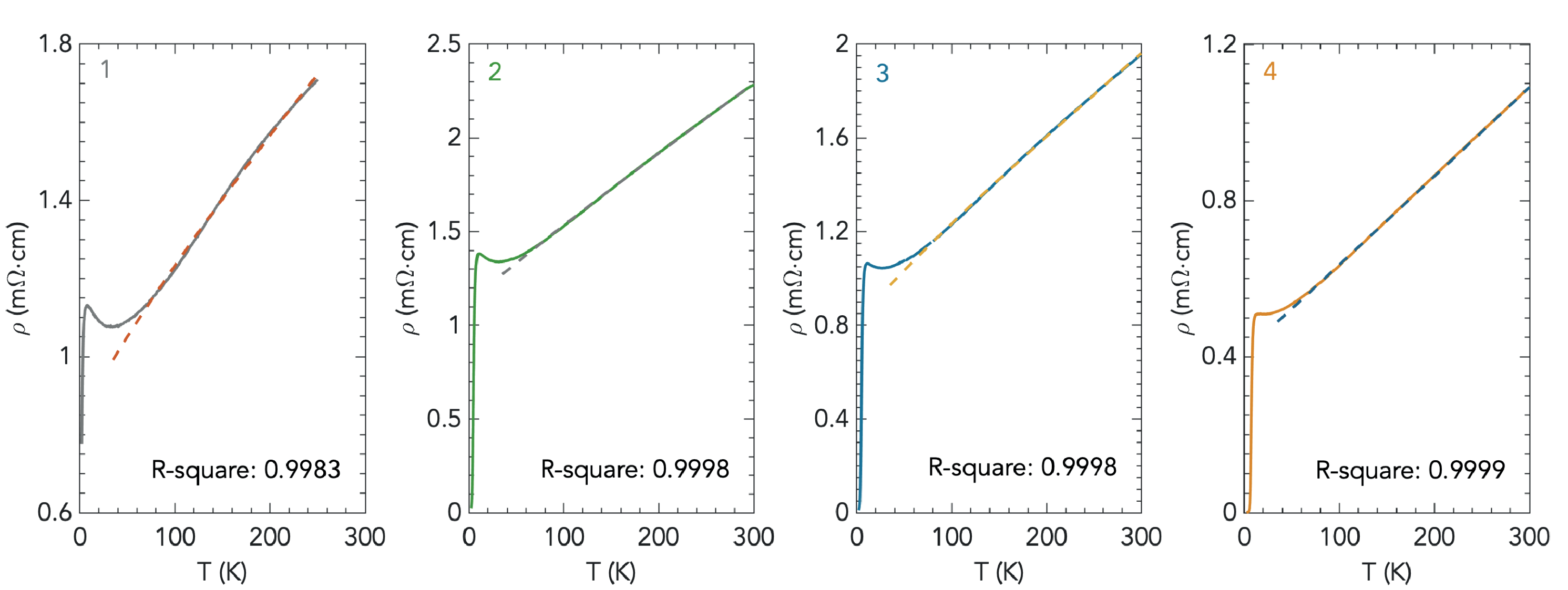} \caption{\label{Fig_linear_fits} {Least-square fitting of the experimental data to $\rho(T) = \rho_{res}+AT^{\alpha}$ in the range from 300 K to 60 K over subsequent reductions.
}}
\end{figure*}

\clearpage

\subsection{Reductions beyond the linear-in temperature resistivity of the normal state}

\begin{figure*}[h]
 \includegraphics[keepaspectratio=true, width=\linewidth]{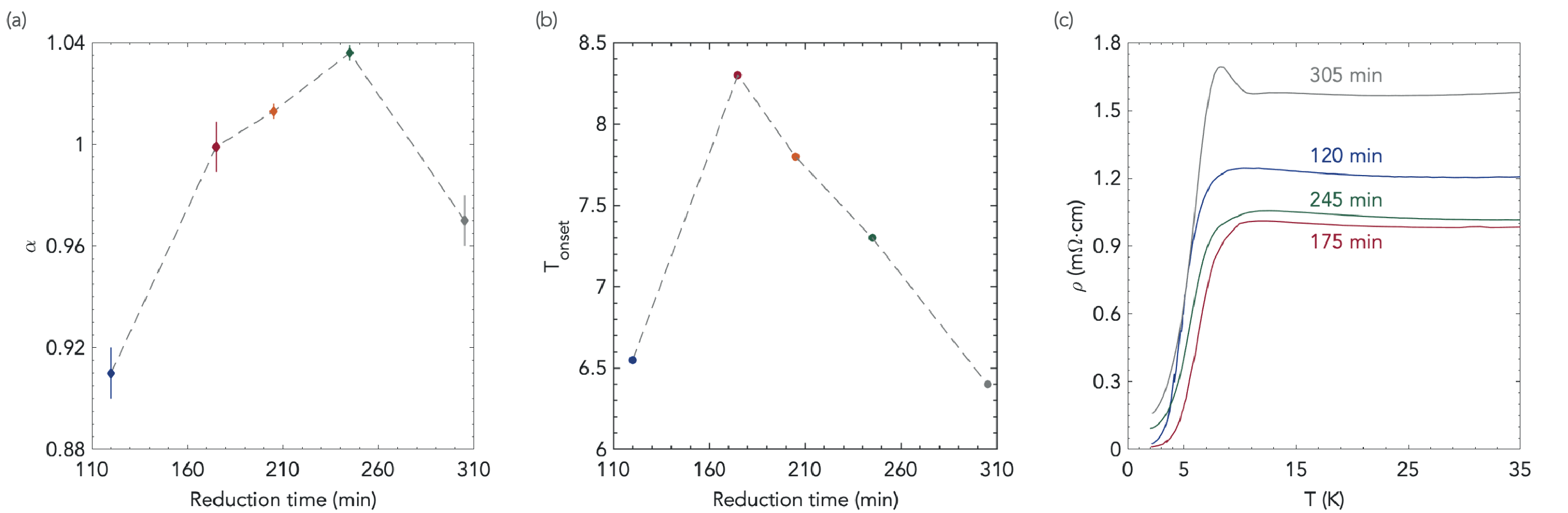} \caption{\label{Fig_reductions_beyond} {(a) Exponent $\alpha$  in $\rho(T)=\rho_{res}+AT^{\alpha}$ of the fitting of the experimental data to $\rho(T) = \rho_{res}+AT^{\alpha}$ in the range from 300 K to 60 K over subsequent reductions. After 245 min of reduction a value of $\alpha \approx 1$ is observed. Reductions beyond that point decrease the onset temperature of the superconducting transition (b), and deteriorate the reduced phase, as shown in the temperature dependence of resistivity (c).
}}
\end{figure*}

\begin{figure*}[h]
 \includegraphics[keepaspectratio=true, width=\linewidth]{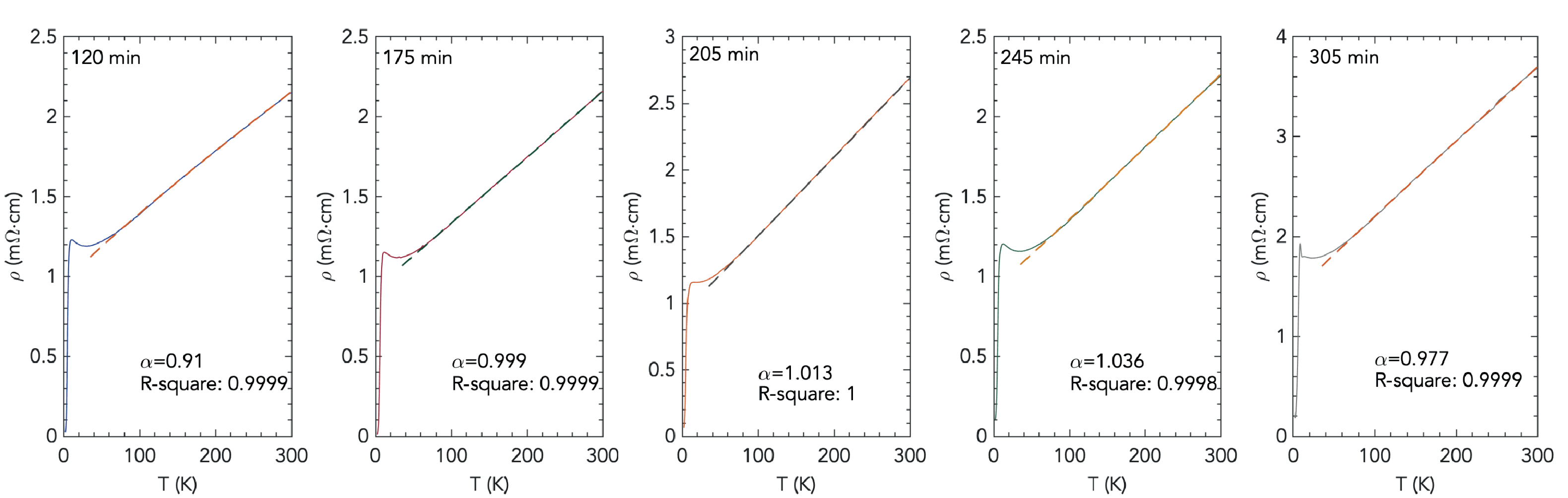} \caption{\label{Fig_fit_reductions_beyond} {Least-square fitting of the experimental data to $\rho(T) = \rho_{res}+AT^{\alpha}$ in the range from 300 K to 60 K over subsequent reductions. The value of the fitting parameter $\alpha$ as a function of the reduction time is shown in Fig.\ref{Fig_reductions_beyond}(a).
}}
\end{figure*}

\clearpage

\subsection{Additional transport characterization of SC PSNO\texorpdfstring{\textsubscript 2}{2} film}

Hall resistivity is obtained by the usual antisymmetrization procedure, carried out between the positive and negative field sweeps in order to cancel out longitudinal resistivity (which is symmetric in the magnetic field).

\begin{figure*}[h]
 \includegraphics[keepaspectratio=true, width=\linewidth]{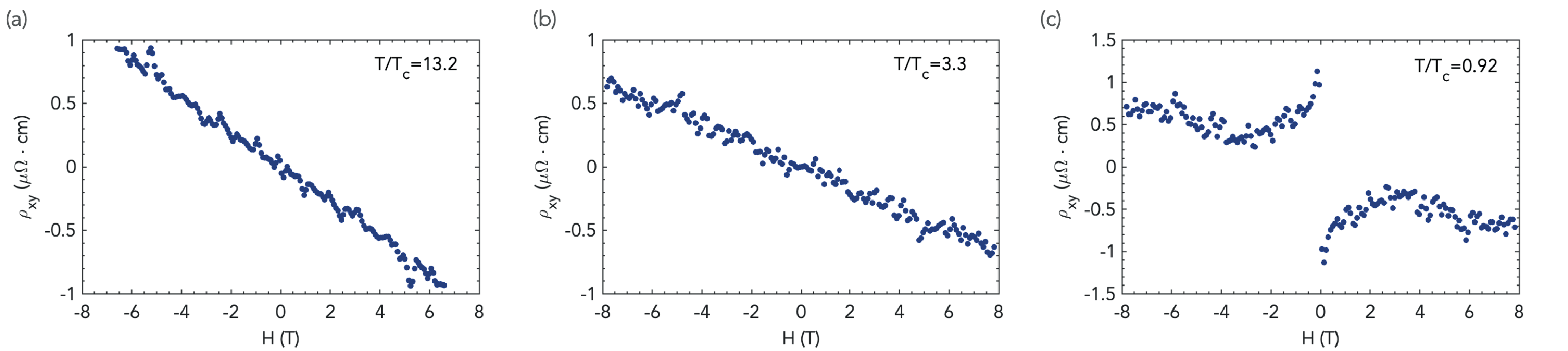} \caption{\label{Fig_Hall_resistance} {Antisymmetrized Hall resistivity as a function of magnetic field at different temperatures for an unpatterned fully reduced SC film ($T_c=7.6\;\mbox{K}$, $\Delta T_c=2.9\;\mbox{K}$). 
}}
\end{figure*}

Hall resistivity versus magnetic field for three different ranges of temperature is shown in Fig.\ref{Fig_Hall_resistance} ((a,b) normal state; (c) mixed state). It leads to a Hall coefficient $R_H=(-1.4 \pm 0.03)\cdot 10^{-3} \mbox{cm}^3\mbox{C}^{-1}$ for $T/T_c=13.2$ (Fig.\ref{Fig_Hall_resistance}(a)) and $R_H=(-8.4 \pm 0.3)\cdot 10^{-4} \mbox{cm}^3\mbox{C}^{-1}$ for $T/T_c=3.3$ (Fig.\ref{Fig_Hall_resistance}(b)).

It is worth noting that typical values of the upper critical field for this system are much higher that the highest magnetic field applied in our experiments (see, for example, Wang et al., Science Advances, 9, 2023, Fig.1(E)).\cite{wang:23}  Thus, the applied magnetic field can enter the superconductor (in the form of quantised flux lines or vortices), which is in mixed state.  Moving vortices lead to dissipation in the mixed state of the superconductor, being associated with the appearance of a non-zero resistivity mixed state Hall effect.  In the critical temperature region (Fig.\ref{Fig_Hall_resistance},(c)) there is a normal-state contribution (linear in H) and a departure from linearity occurs in the magnetic field dependence of Hall resistivity at low fields.  A detailed investigation of the magnetic field dependence of the mixed-state Hall resistivity should be carried out in samples patterned into a Hall bar configuration at various current densities and is beyond the scope of this work.

%A sign change is observed in Hall resistivity at lower temperatures (Fig.\ref{Fig_Hall_resistance},(a)). Whether this sign change is due to the intrinsic property of vortex dynamics or to possible influence of inhomogeneities is unclear.

%\clearpage

\subsection{Scanning transmission electron microscopy of SC PSNO\texorpdfstring{\textsubscript 2}{2} film}

\begin{figure*}[h]
 \includegraphics[keepaspectratio=true, width=0.65\linewidth]{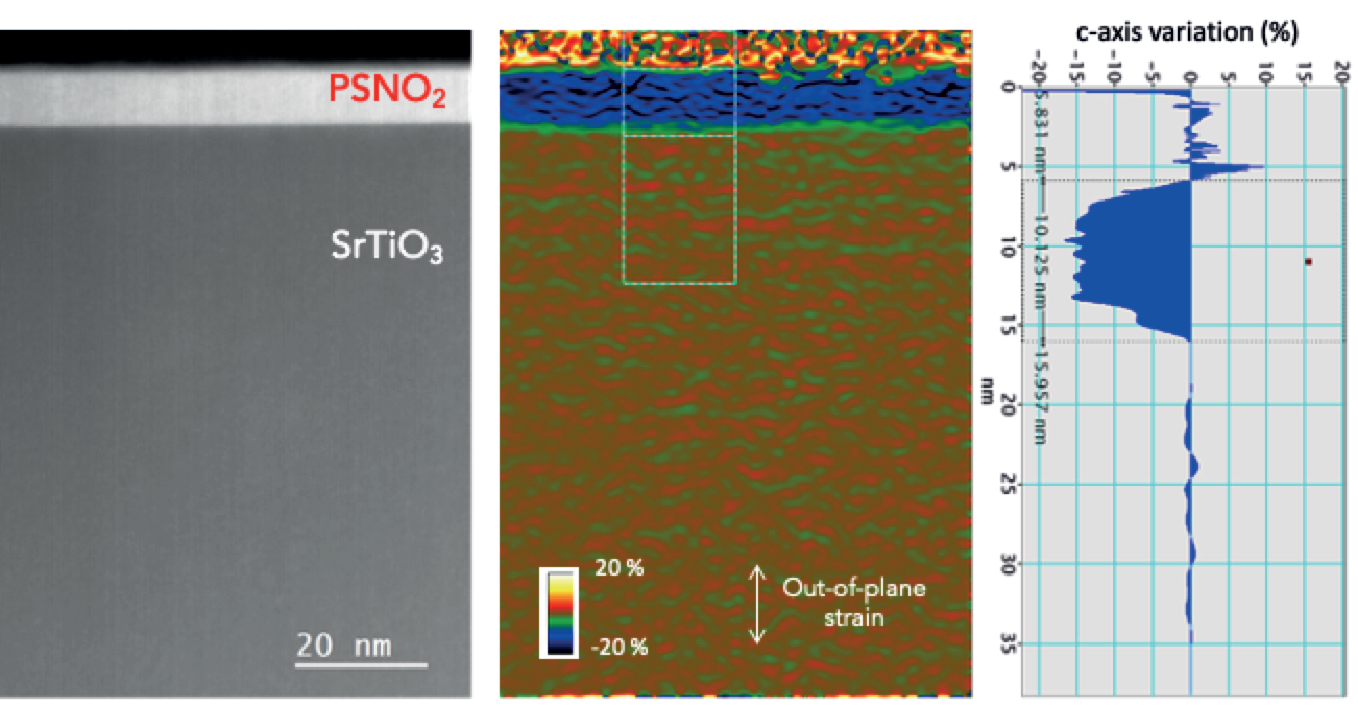} \caption{\label{Fig_TEM_SI_oop_parameter} {Strain map along the out-of-plane direction (center panel) generated from the HAAF-STEM image (left panel) with GPA algorithm. It leads to an estimation of $c\approx 3.32 \; \mbox{\r{A}}$ for the out-of-plane parametter of the IL PSNO$_2$ film (decrease of $\approx$ 15\% related to the lattice parameter of the STO substrate).
}}
\end{figure*}

%\begin{figure*}[h]
% \includegraphics[keepaspectratio=true, width=0.35\linewidth]{Fig_SI_EELS_maps_SC_IL_01} \caption{\label{Fig_SI_EELS_maps_SC_IL} {Elemental EELS maps of Ti L$_{2,3}$, Ni L$_{2,3}$, Pr M$_{4,5}$, and Sr L$_{2,3}$ edges of the PSNO$_2$ film shown in Fig. 5 of the main text.
%}}
%\end{figure*}

%\begin{figure*}[h]
% \includegraphics[keepaspectratio=true, width=0.9\linewidth]{Fig_TEM_SI_interface_termination_01} \caption{\label{Fig_TEM_SI_interface_termination} {Elemental EELS maps of Ti L$_{2,3}$, Ni L$_{2,3}$, Pr M$_{4,5}$, and Sr L$_{2,3}$ edges across the interface of a SC PSNO$_2$ film on STO. The concentration profile reveals an interface Ni-Sr$_{0.8}$Pr$_{0.2}$-Ti on a TiO$_2$-terminated STO substrate.
%}}
%\end{figure*}

\clearpage

\subsection{STEM-EELS element map of PSNO\texorpdfstring{\textsubscript 3}{3}/SrTiO\texorpdfstring{\textsubscript 3}{3} interface}

\begin{figure*}[h]
 \includegraphics[keepaspectratio=true, width=0.65\linewidth]{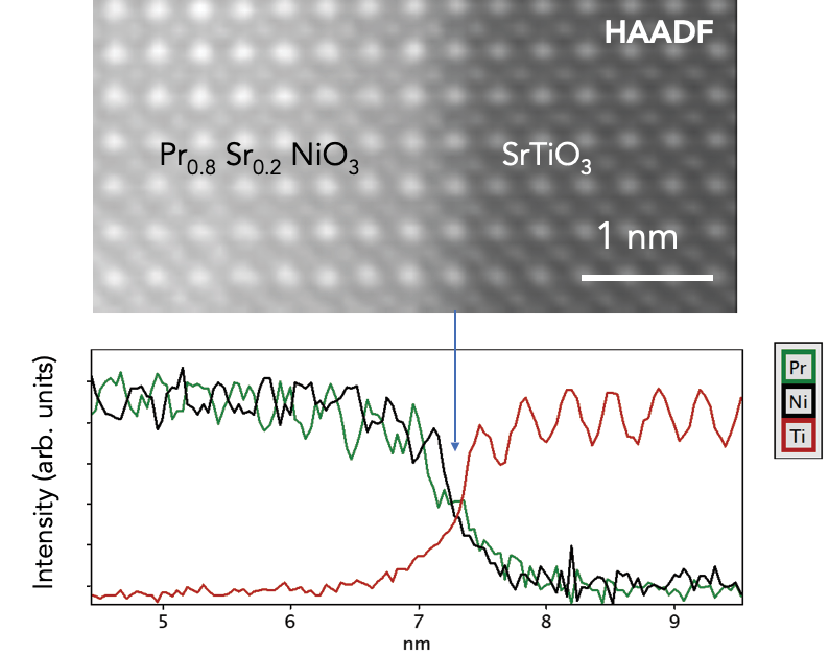} \caption{\label{Fig_interface} {STEM-EELS element map of a PSNO$_3$ thin film on the region near the interface.  HAADF image of the region near the interface with SrTiO$_3$ (top panel) and corresponding element map of Ti-L, Pr-M, and Ni-L edges (bottom panel), showing the absence of the B-site intermixing (Ni/Ti) previously reported at both perovskite and infinite-layer nickelate-substrate interfaces.\cite{goodge:23}
}}
\end{figure*}

%\subsection{Comparison of topochemical reduction on 5 x 5 mm\texorpdfstring{\textsuperscript 2}{2} and 2.5 x 5 mm\texorpdfstring{\textsuperscript 2}{2} samples}

\clearpage

\subsection{Topochemical reduction on uncut samples}

%\begin{figure*}[h]
% \includegraphics[keepaspectratio=true, width=\linewidth]{Fig_S3_04_uncut} \caption{\label{Fig_attemps_on_5x5} {(a) XRD $\theta$-$2\theta$ symmetrical scans on a 5 x 5 mm$^2$ sample as-grown (botton pattern, as reference); (1-1) after a topochemical reduction with CaH$_2$ at 260$^{\circ}$C for 2h45min, and measured immediately after unsealing the ampoule, showing diffraction reflections consistent with a reduced phase; (1-2) measured the next day, showing the oxidised phase, with reflections at the same position than the as-grown sample; (2-2) after a subsequent reduction with CaH$_2$ at 260$^{\circ}$C for additional 4h 15min, measured the second day after unsealing the ampoule.  (b) $\rho(T)$ and (c) $R_H$ of the sample 1-2 (first reduction step, measured carried out 24 hours after the ampoule unsealing), confirmed the reoxidation observed in the XRD pattern in panel (a).  $\rho(T)$ and $R_H$ for the as-grown sample are shown as reference. (d) $\rho(T)$ from the sample (2-1), after second reduction step and measurement carried out right after unsealing the ampoule, showing a SC transition at low temperature consistent with a reduced phase, while the next day the sample has reoxidised as shown in XRD pattern (2-2) in panel (a).  
%}}
%\end{figure*}

\begin{figure*}[!htb]
 \includegraphics[keepaspectratio=true, width=0.55\linewidth]{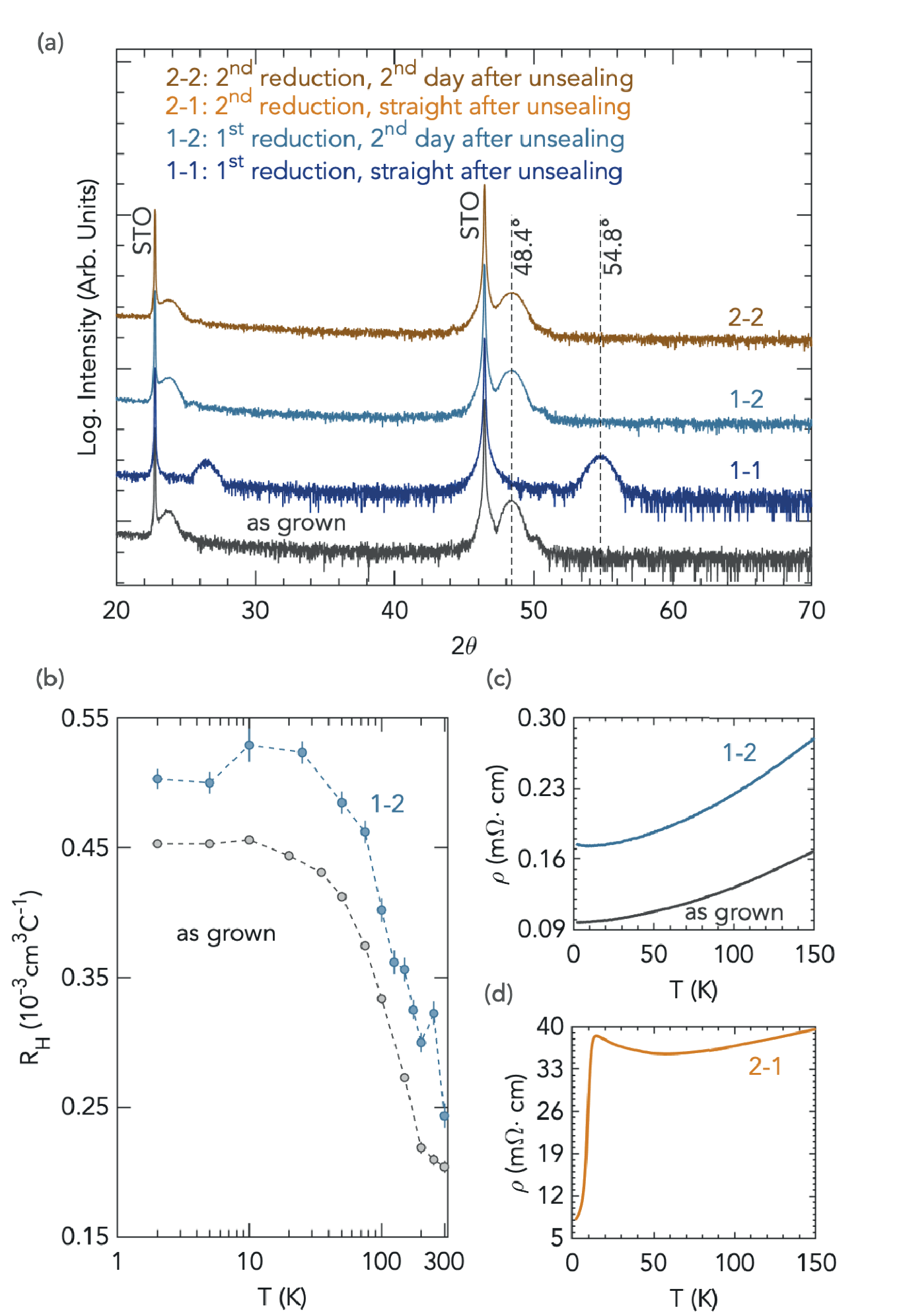} \caption{\label{Fig_attemps_on_5x5} {Topochemical reduction performed on uncut samples of PSNO$_3$.  {\bf{(a)}} XRD $\theta$-$2\theta$ symmetrical scans on a 5 $\times$ 5 mm$^2$ sample as-grown (bottom pattern, as reference); (1-1) after a topochemical reduction with CaH$_2$ at 260$^{\circ}$C for 2h45min, and measured immediately after unsealing the ampoule, showing diffraction reflections consistent with a reduced phase; (1-2) measured the next day, showing the oxidised phase, with reflections at the same position than the as-grown sample; (2-2) after a subsequent reduction with CaH$_2$ at 260$^{\circ}$C for additional 4h 15min, measured the second day after unsealing the ampoule.  {\bf{(b)}} $R_H$ and {\bf{(c)}} $\rho(T)$ of the sample 1-2 (first reduction step, measured 24 hours after the ampoule unsealing), confirmed the reoxidation observed in the XRD pattern in panel{} (a).  $\rho(T)$ and $R_H$ for the as-grown sample are shown as reference. {\bf{(d)}} $\rho(T)$ from the sample (2-1), after second reduction step and measurement carried out right after unsealing the ampoule, showing a SC transition at low temperature consistent with a reduced phase, while the next day the sample has reoxidised as shown in XRD pattern (2-2) in panel ({\textit{a}}).  Error bars indicate the $1\sigma$ uncertainties of the fits.
}}
\end{figure*}

\begin{figure*}[h]
 \includegraphics[keepaspectratio=true, width=0.4\linewidth]{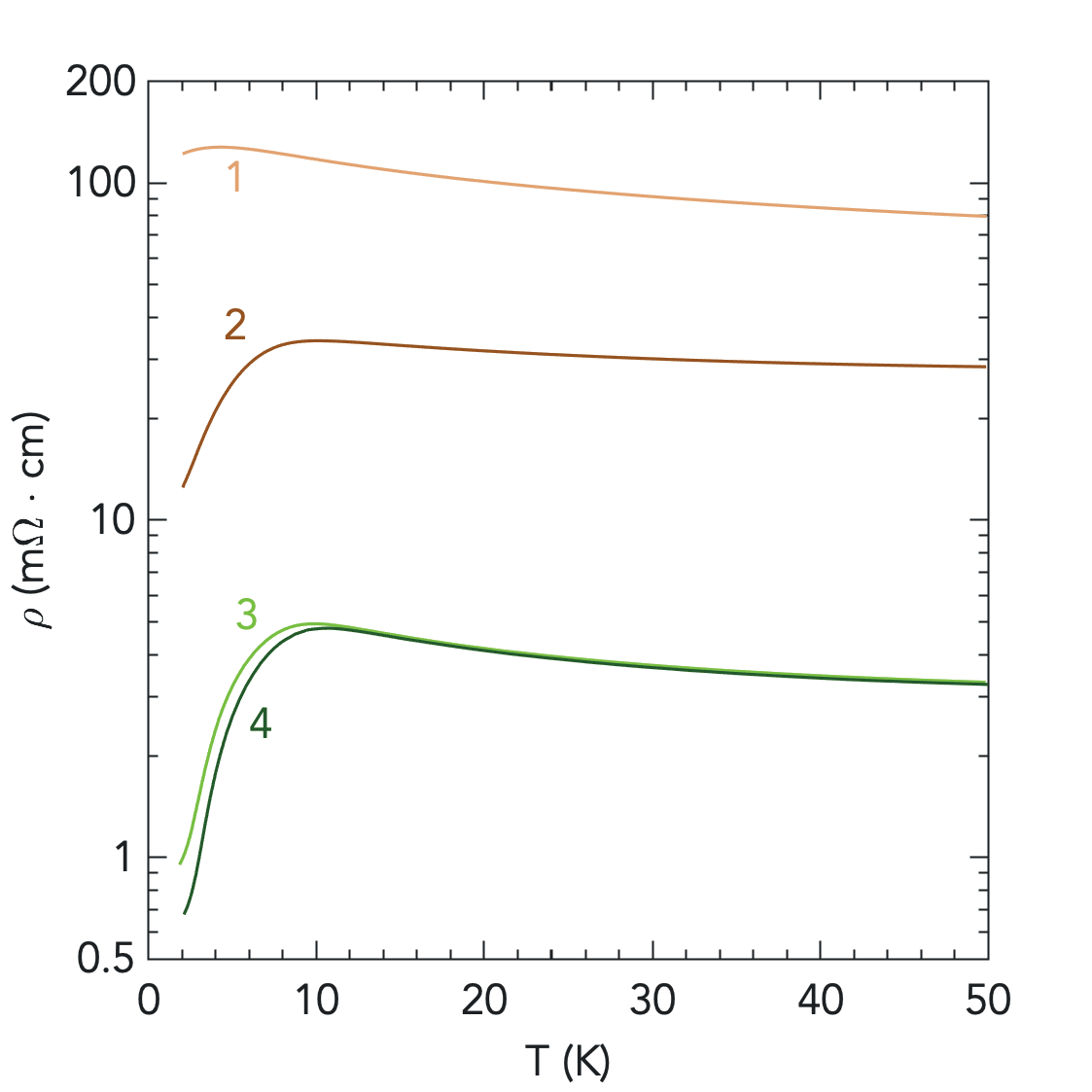} \caption{\label{Fig_cut} {After unproductive reduction attempts, detailed in Fig. \ref{Fig_attemps_on_5x5}, the sample was oxidised in flowing O$_2$ gas (atmospheric pressure) at 680$^{\circ}$C for 12 hours, and cut into four pieces.  Incremental reduction treatments on one of those pieces give rise to a SC transition that improves over successive reduction steps, although a zero-resistance state is not achieved, likely due to some degradation of its crystalline properties after several cycles of reduction/reoxidation that it underwent earlier.
}}
\end{figure*}

\begin{figure*}
 \includegraphics[keepaspectratio=true, width=0.7\linewidth]{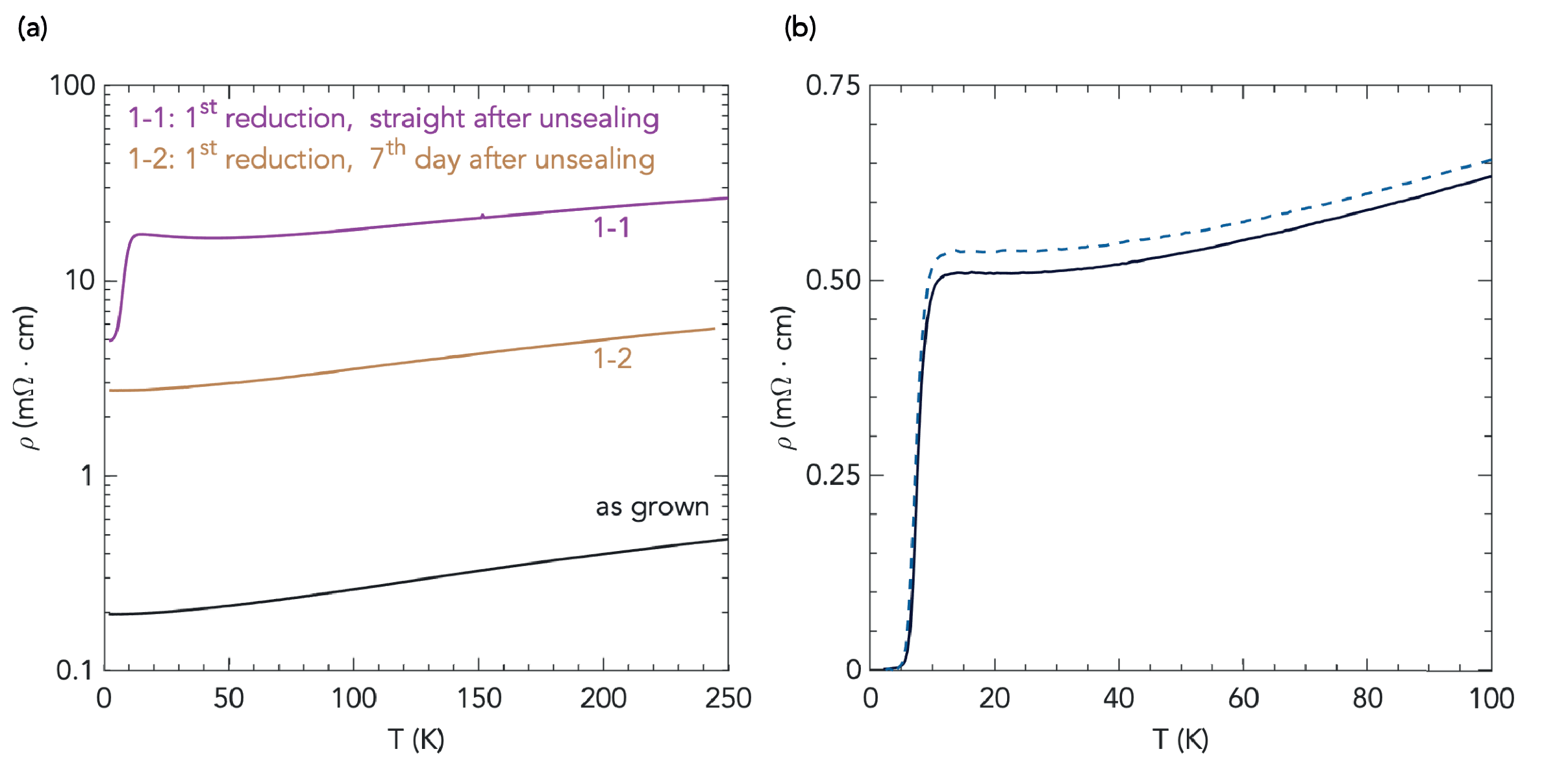} \caption{\label{Fig_unstable_capped_vs_stable} {Reoxidation of the uncut reduced samples is not avoided by an STO capping layer.  {\bf{(a)}} $\rho(T)$ of an uncut film with an STO capping layer immediately after unsealing the ampoule (1-1 plot), and 7 days later (1-2 plot), showing reoxidation.  {\bf{(b)}} No significant change to $\rho(T)$ of SC cut capped samples exposed to the air for up to several weeks is found.  The dashed line plots the resistivity of a cut, capped sample 8 weeks after that plotted in solid line. 
}}
\end{figure*}

Attempts to obtain SC films from uncut samples were always unproductive.  Immediately following unsealing of the ampoule, the uncut film shows XRD patterns or temperature dependence of resistivity typical of a reduced phase, but it readily reoxidizes in less than 24 hours even when stored in a glovebox under nitrogen atmosphere or under vacuum.  This swift reoxidation prevents us from measuring different properties after a reduction step, Fig.\ref{Fig_attemps_on_5x5}  Remarkably, once that sample is cut and the reduction is carried out on one of the resulting pieces, we did not observe reoxidation after successive annealings, as shown in Fig.\ref{Fig_cut}.  Previous works report perovskite films are cut in half before being reduced to the infinite-layer phase,\cite{Li:19,lee:20,lee:23} regardless of the size of the substrate.\cite{fowlie:22}  It is worth noting that reoxidation observed on the uncut reduced samples is not avoided by a STO capping layer, while no reoxidation on capped SC films that have been previously cut was detected within the course of several weeks after reduction (Fig.\ref{Fig_unstable_capped_vs_stable}).  

%This behaviour is strikingly different of that observed in the primary 5 $\times$ 5 mm$^2$ sample under topochemical reduction (Fig. 6, main text).
%5 $\times$ 5 mm$^2$

\clearpage

\subsection{Supplementary details on the topotactic reduction process}

\begin{figure*}[hb]
 \includegraphics[keepaspectratio=true, width=0.65\linewidth]{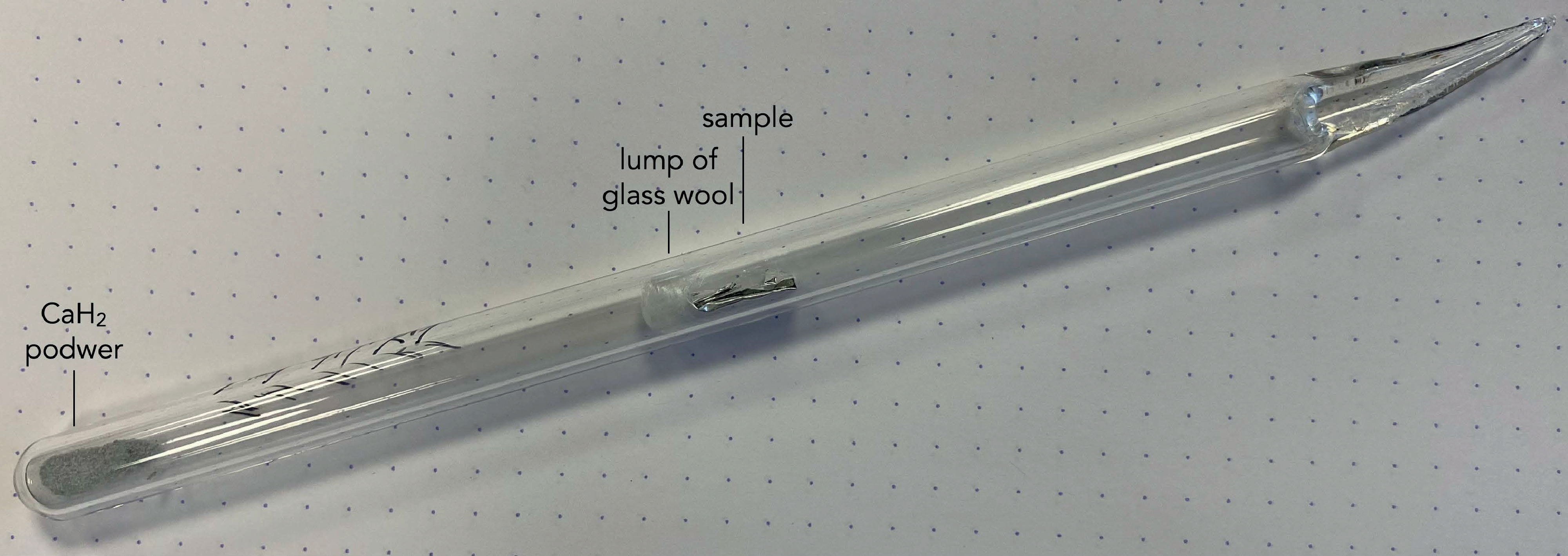} \caption{\label{Fig_ampoule_reduction} {Picture of a sealed tube used for topotactic reduction. The sample is wrapped in aluminum foil and separated from the CaH$_2$ powder by a lump of glass wool. The distance between the sample and the powder is kept constant for increasing reproducibility among reductions. 
}}
\end{figure*}

%\clearpage

\bibliography{supplemental.bib}